\documentclass[trsc,nonblindrev]{informs3} 

\OneAndAHalfSpacedXI 

\usepackage{natbib}
 \bibpunct[, ]{(}{)}{,}{a}{}{,}%
 \def\bibfont{\small}%
\TheoremsNumberedThrough     

\EquationsNumberedThrough    

\MANUSCRIPTNO{XXXX-XXXX} 

\usepackage{amsmath}
\usepackage{algorithm}
\usepackage{algorithmic}
\usepackage{authblk}
\usepackage{xcolor}
\usepackage{lipsum}
\usepackage{amsfonts}

\usepackage{threeparttable}
\usepackage{tabularx}
\usepackage{multirow}
\usepackage{booktabs}
\usepackage{enumerate}
\usepackage{colortbl}
\usepackage{enumitem}
\usepackage{subcaption}
\usepackage{self}
\usepackage{diagbox}

\usepackage{hyperref}
\usepackage{xcolor}
\hypersetup{
	colorlinks,
	linkcolor={red!50!black},
	citecolor={blue!50!black},
	urlcolor={blue!80!black}
}

\graphicspath{{img/}}

\begin{document}


\RUNAUTHOR{Ma and Qian}

\RUNTITLE{PDODE on computational graphs}

\TITLE{Estimating probabilistic dynamic origin-destination demands using multi-day traffic data on computational graphs}

\ARTICLEAUTHORS{%
\AUTHOR{Wei Ma}
\AFF{Department of Civil and Environmental Engineering, The Hong Kong Polytechnic University, Hong Kong SAR\\
The Hong Kong Polytechnic University Shenzhen Research Institute, Shenzhen, Guangdong, China\\
\EMAIL{wei.w.ma@polyu.edu.hk}}
\AUTHOR{Sean Qian}
\AFF{Department of Civil and Environmental Engineering, Carnegie Mellon University, Pittsburgh, PA, USA\\
H. John Heinz III Heinz College, Carnegie Mellon University, Pittsburgh, PA, USA\\
\EMAIL{seanqian@cmu.edu}}
} 

\ABSTRACT{%
System-level decision making in transportation needs to understand day-to-day variation of network flows, which calls for accurate modeling and estimation of probabilistic dynamic travel demand on networks. Most existing studies estimate deterministic dynamic origin-destination (OD) demand, while the day-to-day variation of demand and flow is overlooked.
Estimating probabilistic distributions of dynamic OD demand  is  challenging due to the complexity of the spatio-temporal networks and the computational intensity of the high-dimensional problems.
With the availability of massive traffic data and the emergence of advanced computational methods, this paper develops a data-driven framework that solves the probabilistic dynamic origin-destination demand estimation (PDODE) problem using multi-day data. 
Different statistical distances ({\em e.g.}, $\ell_p$-norm, Wasserstein distance, KL divergence, Bhattacharyya distance) are used and compared to measure the gap between the estimated and the observed traffic conditions, and it is found that 2-Wasserstein distance achieves a balanced accuracy in estimating both mean and standard deviation.
The proposed framework is cast into the computational graph and a reparametrization trick is developed to estimate the mean and standard deviation of the probabilistic dynamic OD demand simultaneously. 
We demonstrate the effectiveness and efficiency of the proposed PDODE framework on both small and real-world networks. 
In particular, it is demonstrated that the proposed PDODE framework can mitigate the overfitting issues by considering the demand variation.
Overall, the developed PDODE framework provides a practical tool for public agencies to understand the sources of demand stochasticity, evaluate day-to-day variation of network flow, and make reliable decisions for intelligent transportation systems.
}%


\HISTORY{Submitted: \today}

\maketitle

%

\section{Introduction}
The spatio-temporal traffic origin-destination demand is a critical component to dynamic system modeling for transport operation and management. For decades, dynamic traffic demand is deterministically modeled as it is translated into deterministic link/path flow and travel cost.  Recent studies on transportation network uncertainty and reliability indicate that the variation of traffic demand also has equally large economic and environmental regional impacts \citep{mahmassani2014incorporating}. However, traffic demand and flow variation from time to time ({\em e.g.}, morning versus afternoon, today versus yesterday) cannot be captured in those deterministic models. In addition, the multi-day large-scale data cannot be characterized by deterministic traffic models. It is essential to consider the flow variation and understand its causes for system-wide decision making in the real-world. 
Therefore, modeling and estimating the stochasticity of traffic demand, namely its spatial-temporal correlation/variation, is a real need for public agencies and decision-makers.

In view of this, this paper addresses a fundamental problem to estimate the probabilistic dynamic origin-destination demand (PDOD) on general road networks. The reasons for estimating the PDOD instead of the deterministic dynamic OD demand (DDOD) are four-fold: 1) PDOD enables the modeling of system variation \citep{han2018stochastic}, and hence the corresponding traffic model is more reliable; 2) later we will show that there is a theoretical bias when using the deterministic dynamic OD demand estimation (DDODE) framework with stochastic traffic flow; 2) the probabilistic dynamic OD estimation (PDODE) framework makes full use of multi-day traffic data, and the confidence level of the estimated PDOD can be quantified. In particular, the confidence in estimation accuracy increases when the number of data increases in the PDODE framework; 4) the estimated PDOD facilitates public agencies to operate and manage the stochastic complex road networks more robustly \citep{jin2019behavior}.

Before focusing on the PDODE problem, we first review the large body of literature for DDODE problems, then their extensions to PDODE problems are discussed.
The DDODE problem is originally proposed and solved through a generalized least square (GLS) formulation by assuming the networks are not congested and travelers' behaviors ({\em e.g.} route choice, departure time choice) are exogenous \citep{cascetta1993dynamic}. On congested networks, travelers' behaviors need to be considered endogenously. A bi-level formulation is then proposed on top of the GLS formulation, in which the upper-level problem solves for the GLS formulation with fixed travelers' behaviors and the lower-level problem updates the travelers' behaviors \citep{tavana2001internally}. Readers are referred to more details on the bi-level formulation from \citet{nguyen1977estimating, leblanc1982selection, fisk1989trip, yang1992estimation, florian1995coordinate, jha2004development,nie2008variational}. The DDODE problem can also be solved with real-time data feeds from and for ATIS/ATMS applications, and state-space models are usually adopted to estimate the OD demand on a rolling basis \citep{bierlaire2004efficient, zhou2007structural, ashok2000alternative}. Another interesting trend is that emerging data sources are becoming available to estimate OD demand directly, which include automatic vehicle identification data \citep{cao2021day}, mobile phone data \citep{bachir2019inferring}, Bluetooth data \citep{cipriani2021traffic}, GPS trajectories \citep{ros2022practical}, and satellite images \citep{kaack2019truck}. Unlike static networks \citep{wu2018hierarchical, waller2021rapidex}, an universal framework that can integrate multi-source data is still lacking for dynamic networks.

Solution algorithms to the DDODE problem can be categorized into two types: 1) meta-heuristic methods; 2) gradient-based methods.
Though meta-heuristics methods might be able to search for the global optimal, most studies only handle small networks with low-dimensional OD demand \citep{patil2022methods}. In contrast, gradient-based methods can be applied to large-scale networks without exploiting computational resources.
The performance of gradient-based methods depends on how to accurately evaluate the gradient of the GLS formulation. \citet{balakrishna2008time, cipriani2011gradient} adopt the stochastic perturbation simultaneous approximation (SPSA) framework to approximate the gradients. \citet{lee2009new, vaze2009calibration, ben2012dynamic, lu2015enhanced, tympakianaki2015c, antoniou2015w, oh2019demand, qurashi2019pc} further enhance the SPSA-based methods. \citet{lu2013dynamic} discuss to evaluate the gradients of dynamic OD under congested networks. \citet{flotterod2011bayesian, yu2021bayesian} derives the gradient of OD demand in a Bayesian inference framework. \citet{osorio2019dynamic, osorio2019high, patwary2021metamodel, dantsuji2022novel} develop a meta-model to approximate the gradients of dynamic OD demand through linear models. Recently, \citet{wu2018hierarchical, ma2019estimating} propose a novel approach to evaluate the gradient of OD demand efficiently through the computational graph approach. 

A few studies have explored the possibilities of estimating PDOD, and this problem turns out to be much more challenging than the DDODE problem. As far as we know, all the existing studies related to the probabilistic OD demand focus on static networks. For example, a statistical inference framework with Markov Chain Monte Carlo (MCMC) algorithm is proposed to estimate the probabilistic OD demand \citep{hazelton2008statistical}. The GLS formulation is also extended to consider the variance/covariance matrices in order to estimate the probabilistic OD demand \citep{shao2014estimation, shao2015estimation}. \citet{ma2018statistical} estimate the probabilistic OD demand under statistical traffic equilibrium using Maximum Likelihood Estimators (MLE). Recently, \citet{yang2019estimating} adopt the Generalized Method of Moment (GMM) methods to estimate the parameters of probability distributions of OD demand. 

Estimating the probabilistic dynamic OD  demand  (PDOD) is challenging, and the reasons are three-fold: 1) PDODE problem requires modeling the dynamic traffic networks in the probabilistic space, hence a number of existing models need to be adapted or re-formulated \citep{shao2006reliability, nakayama2014consistent, watling2015stochastic, ma2017variance}; 2) estimating the probabilistic OD demand is an under-determined problem, and the problem dimension of PDODE is much higher than that for DDODE \citep{shao2015estimation,ma2018statistical,yang2019estimating}; 3) solving PDODE problem is more computationally intensive than solving the DDODE problem, and hence new approaches need to be developed to improve the efficiency of the solution algorithm \citep{flotterod2017search, ma2018estimating, shen2019spatial}.

In both PDODE and DDODE formulations,  travelers' behaviors are modeled through the dynamic traffic assignment (DTA) models. Two major types of DTA models are Dynamic User Equilibrium (DUE) models and Dynamic System Optimal (DSO) models. The DUE models search for the user optimal traffic conditions such that all travelers in the same OD pair have the same utilities \citep{mahmassani1984dynamic, nie2010solving}; DSO models solve the system optimal in which the total disutility is minimized \citep{shen2007path, qian2012system, ma2014continuous}. Most DTA models rely on the Dynamic Network Loading (DNL) models on general networks \citep{ma2008polymorphic}, and the DNL models simulate all the vehicle trajectories and spatio-temporal traffic conditions given the origin-destination (OD) demand and fixed travelers' behaviors.

One noteworthy observation is that many studies have shown great potential in improving the solution efficiency by casting network modeling formulations into computational graphs \citep{wu2018hierarchical, ma2019estimating, sun2019analyzing, zhang2021network, kim2021computational}. The advantages of using computational graphs for PDODE problem lies in that, the computational graph shares similarities with deep neural networks from many perspectives. Hence a number of state-of-art techniques, which are previously developed to enhance the efficiency for solving neural networks, can be directly used for solving the PDODE problem. These techniques include, but are not limited to, adaptive gradient-based methods, dropout \citep{srivastava2014dropout}, GPU acceleration, multi-processing \citep{zinkevich2009slow}. Some of the techniques have been examined in our previous study, and the experimental results demonstrate great potential on large-scale networks \citep{ma2019estimating}. Additionally, multi-source data can be seamlessly integrated into the computational graph to estimate the OD demand.

\begin{figure}[h]
	\centering
	\includegraphics[width=0.85\linewidth]{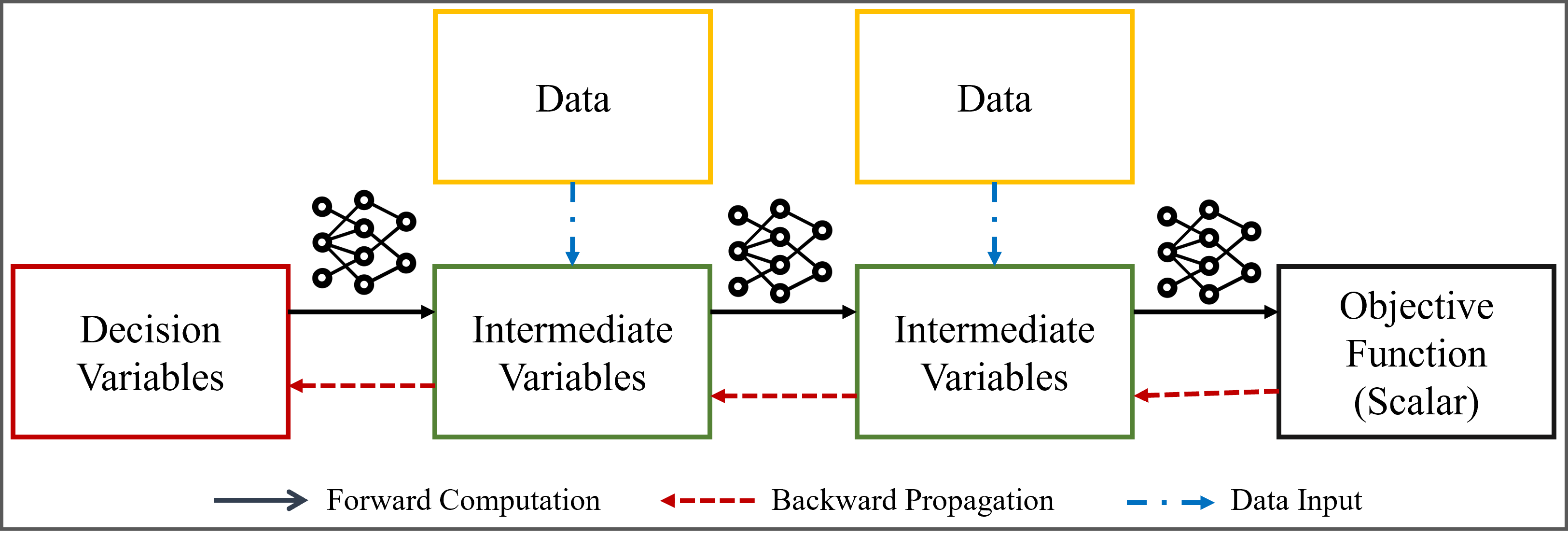}
	\caption{\footnotesize{An illustrative figure for computation graphs.}}
	\label{fig:cg}
\end{figure}

The success of computational graphs advocates the development of the end-to-end framework, and this paper inherits this idea to estimate the mean and standard deviation of PDOD simultaneously. Ideally, the computational graph involves the variables that will be estimated (decision variables, {\em e.g.}, mean and standard deviation of PDOD), intermediate variables ({\em e.g.}, path/link flow distributions), and observed data ({\em e.g.}, observed traffic volumes), and it finally computes the objective function as a scalar. Among different variables, all the neural network operations can be employed. The chain rule and back-propagation can be adopted to update all the variables on the computational graph. This process is also known as  differential programming \citep{jax2018github}. As some of the variables in the computational graph contain physical meaning, we view the computational graph as a powerful tool to integrate  data-driven approaches and domain-oriented knowledge. An overview of a computation graph is presented in Figure~\ref{fig:cg}. 

In this paper, we develop a data-driven framework that solves the probabilistic dynamic OD demand estimation (PDODE) problem using multi-day traffic data on general networks. The proposed framework rigorously formulates the PDODE problem on computational graphs, and different statistical distances ({\em e.g.}, $\ell_p$-norm, Wasserstein distance, KL divergence, Bhattacharyya distance) are used and compared for the objective function. 
The closest studies to this paper are those of \citet{wu2018hierarchical, ma2019estimating}, which construct the computational graphs for static and dynamic OD demand estimation, respectively. This paper extends the usage of computation graphs to solve PDODE problems. The main contributions of this paper are summarized as follows:
\begin{enumerate}[label=\arabic*)]
	\item We illustrate the potential bias in the DDODE framework when dynamic OD demands are stochastic.
	\item We rigorously formulate the probabilistic dynamic OD estimation (PDODE) problem, and different statistical distances are compared for the objective function. It is found that $\ell_1$ and $\ell_2$ norms have advantages in estimating the mean and standard deviation of the PDOD, respectively, and the 2-Wasserstein distance achieves a balanced accuracy in estimating both mean and standard deviation. 
	\item The PDODE formulation is vectorized and cast into a computational graph, and a reparameterization trick is first time developed to estimate the mean and standard deviation of the PDOD simultaneously using adaptive gradient-based methods.
	\item We examine the proposed PDODE framework on a large-scale network to demonstrate its effectiveness and computational efficiency.
\end{enumerate}

The remainder of this paper as organized as follows. Section~\ref{sec:example} illustrates the necessity of the PDODE, and section~\ref{sec:model} presents the proposed model formulation and casts the formulation into computational graphs. Section~\ref{sec:solution} proposes a novel solution algorithm with a reparameterization trick. Numerical experiments on both small and large networks are conducted in section~\ref{sec:experiment}. Finally, conclusions and future research are summarized in section \ref{sec:con}.

\section{An Illustrative Example}
\label{sec:example}
\begin{figure}[h]
	\centering
	\includegraphics[width=0.85\linewidth]{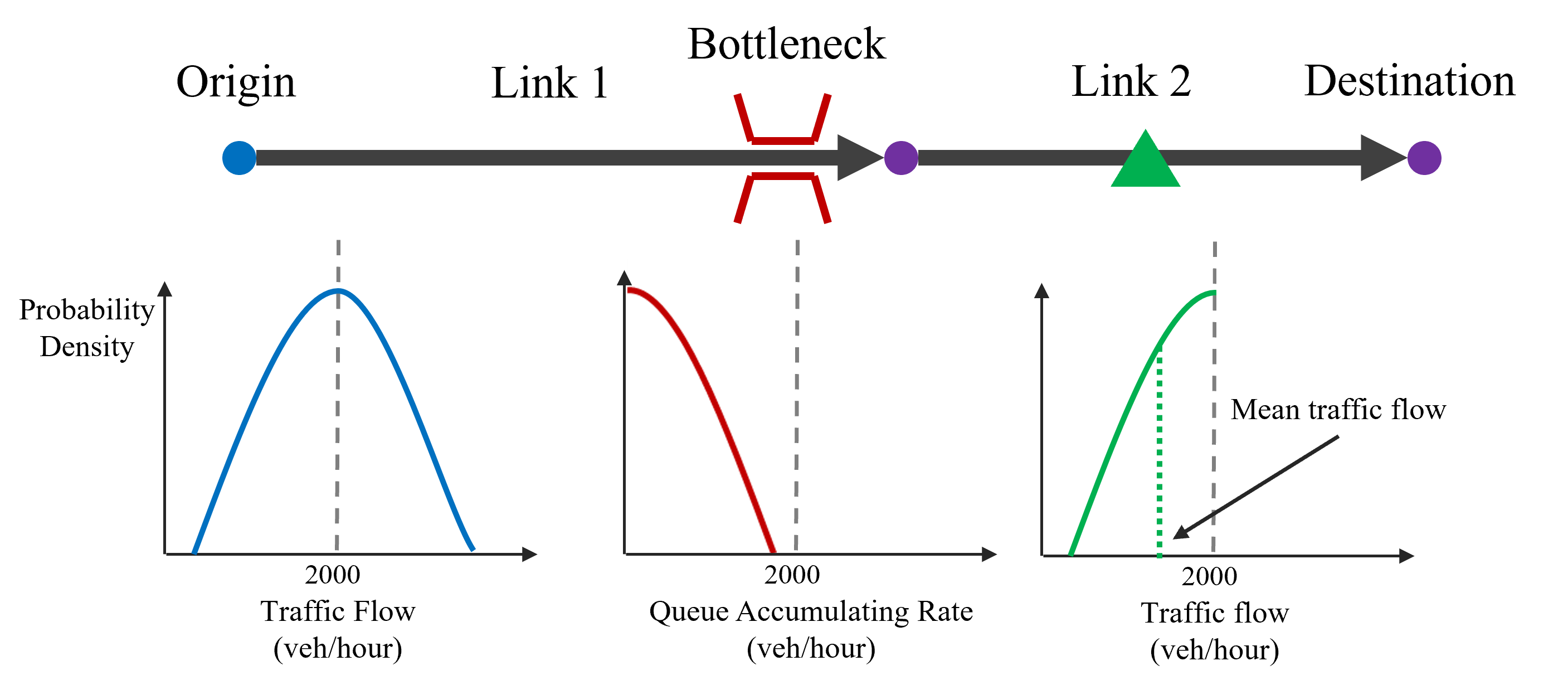}
	\caption{\footnotesize{A simple network with a bottleneck.}}
	\label{fig:bottle}
\end{figure}

To illustrate the necessity of considering the demand variation when the traffic flow is stochastic, we show that the DDODE framework can under-estimate the DDOD (mean of PDOD) when traffic flow is stochastic. Considering a simple 2-link network with a bottleneck, as shown in Figure~\ref{fig:bottle}, the capacity of the bottleneck is 2000vehicles/hour, and the incoming flow follows a Gaussian distribution with mean 2000vehicles/hour. Due to the limited bottleneck capacity, we can compute the probability density functions (PDFs) of the queue accumulative rate and flow rate on the downstream link, as shown in Figure~\ref{fig:bottle}.

Suppose link 2 is installed with a loop detector, we aim to estimate the OD demand from  Origin to Destination. If the DDODE method is used, we ignore the variation in the traffic flow observation on link 2, and the mean traffic flow is used, which is below 2000vehicles/hour. Therefore, the estimated OD demand will be less than 2000vehicles/hour. One can see the demand is under-estimated, and the bias is due to the ignorance of the flow variation. In contrast, the flow variation is considered in our proposed model. By matching the PDF of the observed traffic flow, the distribution of the OD demand can be estimated in an unbiased manner if the model specifications  of the OD demand is correct. Overall, considering flow variation could improve the estimation accuracy of traffic demand, which motivates the development of the PDODE framework.

\section{Model}
\label{sec:model}
In this section, we first present model assumptions. Then important components of the probabilistic traffic dynamics on general networks, which include PDOD, route choice models, and network loading models, are discussed. The PDODE problem is formulated and cast to a vectorized representation using random vectors. Lastly, we propose different statistical distances as the objective function.

\subsection{Assumptions}
Let $Q_{rs}^h$ represent the dynamic traffic demand (number of travelers departing to travel) from OD pair $r$ to $s$ in time interval $h$, where $r \in R, s\in S$ and $h \in H$. $R$ is the set of origins, $S$ is the set of destinations and $H$ is the set of all time intervals. 
\begin{assumption}
	\label{as:mvn}
	The probabilistic dynamic OD demand (PDOD) follows multivariate Gaussian distribution with diagonal covariance matrix. We also assume that $Q_{rs}^h$ is bounded such that $Q_{rs}^h \geq 0$ for the sake of simplification. Readers are referred to \citet{nakayama2016effect} for more discussions about the assumptions of bounded Gaussian distribution of OD demand.
\end{assumption}

\begin{assumption}
	The dynamic traffic flows, including OD demand, path flow, and link flow, are infinitesimal. Therefore, the variation of travelers' choice is not considered \citep{ma2017variance}.
\end{assumption}

\subsection{Modeling the probabilistic network dynamics}
We present different components and their relationship on a probabilistic and dynamic network.
\subsubsection{Probabilistic dynamic OD demand} 
The dynamic OD demand $Q_{rs}^h$ is an univariate random variable, and it can be decomposed into two parts, as shown in Equation~\ref{eq:od}.
\begin{eqnarray}
	\label{eq:od}
	Q_{rs}^h = q_{rs}^h + \varepsilon_{rs}^h
\end{eqnarray}
where $q_{rs}^h$ is the mean OD demand for OD pair $rs$ in time interval $h$ and it is a deterministic scalar; while $\varepsilon_{rs}^h$ represents the randomness of OD demand. Based on Assumption~\ref{as:mvn}, $\varepsilon_{rs}^h$  follows the zero-mean Gaussian distribution, as presented in Equation~\ref{eq:random}.

\begin{eqnarray}
	\label{eq:random}
	\varepsilon_{rs}^h \sim \mathcal{N}\left(0, \left(\sigma_{rs}^h\right)^2  \right)
\end{eqnarray}
where $\sigma_{rs}^h$ is the standard deviation of $Q_{rs}^h$, and $\mathcal{N}(\cdot, \cdot)$ represents the Gaussian distribution.

\subsubsection{Travelers' Route Choice}
To model the travelers' route choice behaviors, we define the time-dependent route choice portion $p_{rs}^{kh}$ such that it distributes OD demand $Q_{rs}^{h}$ to path flow $F_{rs}^{kh}$ by Equation~\ref{eq:ODpath}.
\begin{eqnarray}
	\label{eq:ODpath}
	F_{rs}^{kh} = p_{rs}^{kh} Q_{rs}^h 
\end{eqnarray}
where $F_{rs}^{kh}$ is the path flow (number of travelers departing to travel along a path) for $k$th path in OD pair $rs$ in time interval $h$. The route choice portion $p_{rs}^{kh}$ can be determined through a generalized route choice model, as presented in Equation~\ref{eq:gen_choice}.
\begin{eqnarray}
	\label{eq:gen_choice}
	p_{rs}^{kh} = \Psi_{rs}^{kh}\left( \D\left(\{C_{rs}^{kh}\}_{rskh}\right), \D\left(\{T_{a}^h\}_{ah}\right)\right)
\end{eqnarray}
where $\Psi_{rs}^{kh}$ is the generalized route choice model and the operator $\D(\cdot)$ extracts all the parameters in a certain distribution. For example, if $Y\sim \N(\mu, \sigma^2)$, then $\D(Y) = (\mu, \sigma)$. $T_{a}^{h}$ represents the link travel time for link $a$ in time interval $h$, and $C_{rs}^{kh}$ represents the path travel time for $k$th path in OD pair $rs$ departing in time interval $h$. Equation~\ref{eq:gen_choice} indicates that the route choice portions are based on the distributions of link travel time and path travel time. In this paper we use travel time as the disutility function, while any form of disutility can be used as long as it can be simulated by $\Lambda$. The generalized travel time can include roads tolls, left turn penalty, delay at intersessions, travelers' preferences and so on.

\subsubsection{Dynamic network loading} For a dynamic network, the network conditions ({\em i.e.} path travel time, link travel time, delays) are governed by the link/node flow dynamics, which can be modeled through dynamic network loading (DNL) models \citep{ma2008polymorphic}. Let $\Lambda(\cdot)$ represent the DNL model, as presented in Equation~\ref{eq:dnl}.
\begin{eqnarray}
	\label{eq:dnl}
	\left\{T_{a}^{h}, C_{rs}^{kh}, {\rho}_{rs}^{ka}(h, h') \right\} _{r,s,k,a,h, h'} =  \Lambda(\{F_{rs}^{kh}\}_{r,s,k,h})
\end{eqnarray}
where $\rho_{rs}^{ka}(h, h')$ is the dynamic assignment ratio (DAR) which represents the portion of the $k$th path flow departing within time interval $h$ for OD pair $rs$ which arrives at link $a$ within time interval $h'$ \citep{ma2018estimating}. Link $a$ is chosen from the link set $A$, and path $k$ is chosen from the set of all paths for OD $rs$, as represented by $a \in A, k \in K_{rs}$. We remark that $T_{a}^{h}, C_{rs}^{kh}, \rho_{rs}^{ka}(h, h')$ are random variables as they are the function outputs of the random variable $F_{rs}^{kh}$. 

The DNL model $\Lambda$ depicts the network dynamics through traffic flow theory \citep{zhang2013modelling,jin2012link}. Essentially, many existing traffic simulation packages, which include but are not limited to, MATSIM \citep{balmer2009matsim}, Polaris \citep{stolte2002polaris}, BEAM \citep{sheppard2017modeling}, DynaMIT \citep{ben1998dynamit}, DYNASMART \citep{mahmassani1992dynamic}, DTALite \citep{zhou2014dtalite} and MAC-POSTS \citep{qian2016dynamic, CARTRUCK}, can be used as function $\Lambda$. In this paper, MAC-POSTS is used as $\Lambda$.

Furthermore, link flow can be modeled by Equation~\ref{eq:link}.
\begin{eqnarray}
	\label{eq:link}
	X_a^{h'} = \sum_{rs \in K_q} \sum_{k \in K_{rs}} \sum_{h \in H}{\rho}_{rs}^{ka}(h, h') F_{rs}^{kh}
\end{eqnarray}
where $X_a^{h'}$ represents the flow of link $a$ that arrives in time intervel $h'$, and $K_q$ is the set of all OD pairs. 

\subsubsection{Statistical equilibrium}
The route choice proportion $p_{rs}^{kh}$ is a  deterministic variable rather than a random variable, and the reason is that we assume travelers' behaviors are based on the statistical equilibrium originally defined in \citet{ma2017variance}, as presented in Definition~\ref{def:equi}.
\begin{definition}
	\label{def:equi}
	A road network is under statistical equilibrium, if all travelers practice the following behavior: on each day, each traveler from origin $r$ to destination $s$ independently chooses route $k$ with a deterministic probability $p_{rs}^k$. For a sufficient number of days, this choice behavior yields a stabilized distribution of travel costs with parameters $\D\left(\{C_{rs}^{kh}\}_{rskh}\right), \D\left(\{T_{a}^h\}_{ah}\right)$. This stabilized distribution, in turn, results in the deterministic probabilities $p_{rs}^{kh} = \Psi_{rs}^k\left( \D\left(\{C_{rs}^{kh}\}_{rskh}\right), \D\left(\{T_{a}^h\}_{ah}\right)\right)$ where $\psi_{rs}^k(\cdot)$ is a general route choice function.  Mathematically, we say the network is under statistical equilibrium when the random variables $(\{Q_{rs}^h \}_{rsh},\{F_{rs}^{kh}\}_{rskh},  \{X_a^h\}_{ah},  \{C_{rs}^{kh}\}_{rskh}, \{T_a^h\}_{ah})$ are consistent with the Formulation \ref{eq:ODpath}, \ref{eq:gen_choice} and \ref{eq:dnl}.
\end{definition}

The statistical equilibrium indicates that the multi-day traffic conditions are independent and identically distributed, which differs from the assumptions in day-to-day traffic models. Readers are referred to \citet{ma2017variance} for more details. The assumption of statistical equilibrium allows us to estimate the distribution of link/path/OD flow from the observed data, and the empirical covariance matrix of link/path/OD flow can approximate the corresponding true covariance matrix when there is a large number of data.

\subsubsection{Vectorization} 
To simplify notations, all the related variables are vectorized. We set $N = |H|$ and denote the total number of paths as $\Pi = \sum_{rs} |K_{rs}|$, $K=|K_q|$. The vectorized variables are presented in Table~\ref{tab:mcvec}.
\begin{table*}[h]
	\begin{center}
		\caption{Variable vectorization table (R.V.: random variable).}
		\label{tab:mcvec}
		\begin{tabular}{p{3cm}cccccp{4.5cm}}
			\hline
			Variable & R.V. &Scalar & Vector& Dimension & Type & Description\\
			\hline\hline \rule{0pt}{3ex}
			Mean OD flow & No &$q_{rs}^h$ & $\vec{q}$ &$\mathbb{R}^{NK}$ & Dense & $q_{rs}^h$ is placed at entry $(h-1)K + k$\\
			\hline \rule{0pt}{3ex}
			Standard deviation of OD flow & No &$\sigma_{rs}^h$ & $\boldsymbol \sigma$ &$\mathbb{R}^{NK}$ &  Dense & $\sigma_{rs}^h$ is placed at entry $(h-1)K + k$\\
			\hline \rule{0pt}{3ex}
			Randomness of OD flow & Yes &$\varepsilon_{rs}^h$ & $\boldsymbol  \varepsilon$ &$\mathbb{R}^{NK}$ & Dense & $\varepsilon_{rs}^h$ is placed at entry $(h-1)K + k$\\
			\hline \rule{0pt}{3ex}
			OD flow & Yes&$Q_{rs}^h$ & $\vec{Q}$ &$\mathbb{R}^{NK}$ & Dense & $Q_{rs}^h$ is placed at entry $(h-1)K + k$\\
			\hline \rule{0pt}{3ex}
			Path flow & Yes &$F_{rs}^{kh}$ &$\mathbf{F}$ & $\mathbb{R}^{N\Pi}$ & Dense & $F_{rs}^{kh}$ is placed at entry $(h-1)\Pi + k$\\
			\hline \rule{0pt}{3ex}
			Link flow & Yes &$X_{a}^h$ & $\mathbf{X}$ &$\mathbb{R}^{N|A|}$ & Dense & $X_{a}^h$ is placed at entry $(N-1)|A| + k$\\
			\hline \rule{0pt}{3ex}
			Link travel time & Yes&$T_{a}^h$ & $\mathbf{T}$ &$\mathbb{R}^{N|A|}$ & Dense & $T_{a}^h$ is placed at entry $(N-1)|A| + k$\\
			\hline \rule{0pt}{3ex}
			Path travel time & Yes&$C_{rs}^{kh}$ & $\mathbf{C}$ &$\mathbb{R}^{N\Pi}$ & Dense & $C_{rs}^{kh}$ is placed at entry $(N-1)\Pi + k$\\
			\hline \rule{0pt}{3ex}
			DAR matrix & Yes&$\rho_{rs}^{ka}(h, h')$ &$\boldsymbol \rho$ & $\mathbb{R}^{N|A| \times N\Pi}$ & Sparse & $\rho_{rs}^{ka}(h, h')$ is placed at entry $[(h'-1)|A| + a, (h-1)\Pi + k]$\\
			\hline \rule{0pt}{3ex}
			Route choice matrix & No&$p_{rs}^{kh}$ &$\mathbf{p}$ & $\mathbb{R}^{N\Pi \times NK}$ & Sparse & $p_{rs}^{kh}$ is placed at entry $[(h-1)|\Pi| + k, (h-1)K + rs]$\\
			\hline
		\end{tabular}
	\end{center}
\end{table*}

Using the notations presented in Table~\ref{tab:mcvec}, we can rewrite Equation~\ref{eq:od}, \ref{eq:random}, \ref{eq:ODpath}, \ref{eq:gen_choice}, \ref{eq:dnl}, and \ref{eq:link} in Equation~\ref{eq:vec}.
\begin{equation}
	\label{eq:vec}
	\begin{array}{llllll}
		\vspace{5pt}
		\vec{Q} &=& \vec{q} + \boldsymbol \varepsilon\\
		{\boldsymbol \varepsilon} &\sim& \N\left(\vec{0}, {\boldsymbol \sigma}^2\right)\\
		\vec{F} &=& \vec{p}\vec{Q}\\
		\vec{p} &= & \Psi \left( \D \left(\vec{C} \right), \D \left(\vec{T}\right)  \right)&  \\
		\left \{ \vec{C}, \vec{T}, {\boldsymbol\rho} \right\} &= & \Lambda(\vec{F}) & \\
		\vec{X} &=&{\boldsymbol \rho} \vec{F}
	\end{array}
\end{equation}
where ${\boldsymbol \sigma}^2$ denotes the element-wise square for matrix ${\boldsymbol \sigma}$. In the rest of this paper, we will use the vectorized notations for simplicity.

\subsection{Formulating the PDODE problem}
The PDODE problem is formulated in this section. In particular, different objective functions are discussed.
\subsubsection{Objective function} 
\label{sec:obj}
To formulate the PDODE problem, we first define the objective function in the optimization problem. DDODE problem minimizes the gap between the estimated (reproduced) and the observed traffic conditions. The gap is usually measured through $\ell^2$-norm, which is commonly used to measure the distance between two deterministic variables. However, the PDODE problem is formulated in the probabilistic space, and we need to measure the distance between the distributions of the observed traffic conditions and the  estimated (reproduced) traffic conditions. 
To this end, we define a generalized form to measure the observed and estimated distribution of traffic conditions, as presented by Equation~\ref{eq:ere}.

\begin{eqnarray}
	\label{eq:ere}
	\mathcal{L}_0 = \mathcal{M} \left(\tilde{\mathbf{X}}, \mathbf{X}(\mathbf{Q})\right) 
\end{eqnarray}
where $\mathcal{M}$ measures the statistical distance between two distributions, which is defined in Definition~\ref{def:stat}.

\begin{definition}
\label{def:stat}
The statistical distance $\mathcal{M}(\mathbf{X}_1, \mathbf{X}_2)$  is defined as the distance between two random vectors ({\em i.e.}, two probabilistic distributions) $\mathbf{X}_1$ and $\mathbf{X}_2$, and it should satisfy two properties: 1) $\mathcal{M}(\mathbf{X}_1, \mathbf{X}_2) \geq 0, \forall~\mathbf{X}_1, \mathbf{X}_2 $; 2) $\mathcal{M}(\mathbf{X}_1, \mathbf{X}_2) = 0  \iff \mathbf{X}_1 = \mathbf{X}_2$. 
\end{definition}

The statistical distance may not be symmetric with respect to $\mathbf{X}_1$ and $\mathbf{X}_2$, and hence it may not be viewed as a metric. Various statistical distances can be used for $\mathcal{M}$, and we review existing literature to list out some commonly used distances that have explicit form for Gaussian distributions. We further simplify the notation $\mathbf{X}(\mathbf{Q})$ to $\mathbf{X}$, and we assume $\tilde{\mathbf{X}} \sim \mathcal{N}\left( \tilde{\mathbf{x}},  \Sigma_{\tilde{\mathbf{X}}} \right)$ and $\mathbf{X} \sim \mathcal{N}\left( \mathbf{x},  \Sigma_{\mathbf{X}} \right)$, then  different statistical distances can be computed as follows.
\begin{itemize}
	\item $\ell_p$-norm on distribution parameters: this metric directly compare the $\ell_p$-norm of the mean vector and covariance matrix, which can be written as:
	$$\|\tilde{\mathbf{x}} - \mathbf{x}\|_p + \| \Sigma_{\tilde{\mathbf{X}}} - \Sigma_{\mathbf{X}}\|_p $$
	\item Wasserstein distance: the 2-Wasserstein distance has close-form for Gaussian distributions, and $\mathcal{M}\left(\tilde{\mathbf{X}}, \mathbf{X}\right)$ can be written as:
	$$\|\tilde{\mathbf{x}} - \mathbf{x}\|_2^2 + \text{Tr}\left( \Sigma_{\tilde{\mathbf{X}}} + \Sigma_{\mathbf{X}} -2 \left(\Sigma_{\tilde{\mathbf{X}}}^{1/2}  \Sigma_{\mathbf{X}} \Sigma_{\tilde{\mathbf{X}}}^{1/2}  \right)^{1/2} \right)$$
	\item Kullback–Leibler (KL) divergence: also known as relative entropy. KL divergence is not symmetric, and we choose the forward KL divergence to avoid taking inverse of $\Sigma_{\mathbf{X}}$, which can be written as 
	$$\frac{1}{2} \left[ \log \frac{|\Sigma_{\tilde{\mathbf{X}}}|}{|\Sigma_{\mathbf{X}}|} +  (\tilde{\mathbf{x}} - \mathbf{x})^T\Sigma_{\tilde{\mathbf{X}}}^{-1} (\tilde{\mathbf{x}} - \mathbf{x})  + \text{Tr}\left(\Sigma_{\tilde{\mathbf{X}}}^{-1} \Sigma_{\mathbf{X}}\right) \right]. $$
	We note that KL divergence can be further extended to Jensen–Shannon (JS) divergence, while it requires to take the inverse of $\Sigma_{\mathbf{X}}$, so we will not consider it in this study.
	\item Bhattacharyya distance: we set $\Sigma = \frac{\Sigma_{\tilde{\mathbf{X}}} + \Sigma_{\mathbf{X}}}{2}$, then $\mathcal{M}\left(\tilde{\mathbf{X}}, \mathbf{X}\right)$ can be written as:
	$$\frac{1}{8}(\tilde{\mathbf{x}} - \mathbf{x})^T\Sigma^{-1} (\tilde{\mathbf{x}} - \mathbf{x}) + \frac{1}{2}\ln \frac{|\Sigma|}{\sqrt{|\Sigma_{\tilde{\mathbf{X}}}|  |\Sigma_{\mathbf{X}}|}}$$
\end{itemize}

All the above statistical distances satisfy Definition~\ref{def:stat}, and they are continuous with respect to the distribution parameters. More importantly, all the statistical distances are differentiable, as each of the used operation is differentiable and the auto-differentiation techniques can be used to derive the overall gradient of the statistical distances with respect to the distribution parameters \citep{speelpenning1980compiling}. Theoretically, all the above distances can be used as the objective function in PDODE, while we will show in the numerical experiments that their performance can be drastically different.


\subsubsection{PDODE formulation}

To simulate the availability of multi-day traffic data, we assume that $I$ day's traffic counts data are collected, and $\tilde{\mathbf{X}}^{(i)}$ is the $i$th observed link flow data, $i= 1, 2, \cdots, I$, and $\tilde{\mathbf{X}}^{(i)}$ iid follows the distribution of $\tilde{\mathbf{X}}$.
Because the actual distributions of $\tilde{\mathbf{X}}$ and $\mathbf{X}$ are unknown, we use the Monte-Carlo approximation to approximate $\mathcal{L}_0$, as presented in Equation~\ref{eq:ereap}.
\begin{eqnarray}
	\label{eq:ereap}
	\mathcal{L} &=& \mathbb{E}_{\left({\boldsymbol \alpha}, {\boldsymbol \beta}\right) \sim  
		{\tilde{\mathbf{X}}}^{\bigotimes I} \bigotimes {\mathbf{X}}^{\bigotimes L}} \mathcal{M} \left(\boldsymbol \alpha, \boldsymbol \beta\right) \nonumber\\
	&=&\frac{1}{IL}\sum_{i=1}^I \sum_{l=1}^{L} \mathcal{M} \left(\tilde{\mathbf{X}}^{(i)}, \mathbf{X}^{(l)}\right) \label{eq:L}
\end{eqnarray}
where $I, L$ are the number of samples from distributions of $\tilde{\mathbf{X}}$ and $\mathbf{X}$, respectively, and $\tilde{\mathbf{X}}^{(i)}, \mathbf{X}^{(l)}$ are the sampled distributions of $\tilde{\mathbf{X}}$ and $\mathbf{X}$, respectively.
By the law of large numbers (LLN), $\mathcal{L}$ converges to $\mathcal{L}_0$ when $I,L \to \infty$.

Combining the constraints in Equation~\ref{eq:vec} and the objective function in Equation~\ref{eq:L}, now we are ready to formulate the PDODE problem in Formulation~\ref{eq:pdode1}.

\begin{equation}
	\label{eq:pdode1}
	\begin{array}{rrcllll}
		\vspace{5pt}
		\displaystyle \min_{\vec{q}, {\boldsymbol \sigma}} & \multicolumn{4}{l}{\displaystyle    \frac{1}{IL}\sum_{i=1}^I \sum_{l=1}^{L} \mathcal{M} \left(\tilde{\mathbf{X}}^{(i)}, \mathbf{X}^{(l)}\right)} &\\
		\textrm{s.t.} & \left \{ \vec{C}^{(l)}, \vec{T}^{(l)}, {\boldsymbol \rho}^{(l)} \ \right\} &= & \Lambda(\vec{F}^{(l)}) & \forall l&\\
		~ & \vec{p}^{(l)} &= & \Psi \left( \D(\vec{C}^{(l)}), \D(\vec{T}^{(l)})  \right)&  \forall l &\\
		~ & \mathbf{Q}^{(l)} &\sim& \mathcal{N}\left(\vec{q}, {\boldsymbol \sigma}^2\right)&\forall l&\\
		~ & \vec{F}^{(l)} & = & \vec{p}^{(l)}\vec{Q}^{(l)} & \forall l&\\
		~ & \mathbf{X}^{(l)} & = &  {\boldsymbol \rho}^{(l)} \vec{F}^{(l)} &\forall l &
	\end{array}
\end{equation}
where $\vec{C}^{(l)}, \vec{T}^{(l)}, \mathbf{X}^{(l)}, \vec{F}^{(l)}, \mathbf{Q}^{(l)}, {\boldsymbol \rho}^{(l)}, \vec{p}^{(l)}$ are the sample distributions of $\vec{C}, \vec{T}, \mathbf{X}, \vec{F}, \mathbf{Q}, {\boldsymbol \rho}, \vec{p}$, respectively. Formulation~\ref{eq:pdode1} searches for the optimal mean and standard deviation of the dynamic OD demand to minimize the statistical distance between the observed and estimated link flow distributions such that the DNL and travelers' behavior models are satisfied.  We note that Formulation~\ref{eq:pdode1} can be extended to include the traffic speed, travel time, and historical OD demand data \citep{ma2019estimating}. It is straightforward to show that Formulation~\ref{eq:pdode1} is always feasible, as long as the sampled PDOD is feasible to the traffic simulator, as presented in Proposition~\ref{prop:fea}.

\begin{proposition}[Feasibility]
\label{prop:fea}
There exist a feasible solution $(\vec{q}, {\boldsymbol \sigma})$ to Formulation~\ref{eq:pdode1} if the non-negative support of the distribution $\mathcal{N}\left(\vec{q}, {\boldsymbol \sigma}^2\right)$ is feasible to the traffic simulator $\Lambda$.
\end{proposition}

To compute $\mathcal{M} \left(\tilde{\mathbf{X}}^{(i)}, \mathbf{X}^{(l)}\right)$, we first characterize the distribution of $\vec{X}^{(l)}$ by $\vec{X}^{(l)} = \Lambda(\vec{p}^{(l)} \vec{Q}^{(l)})$.
Hence the computation of the $\mathcal{M} \left(\tilde{\mathbf{X}}^{(i)}, \mathbf{X}^{(l)}\right)$ is based on the distribution of $\vec{Q}^{(l)}$, as the distribution of $\mathbf{X}^{(l)}$ is obtained from $\mathbf{Q}^{(l)}$. Additionally, the sample distribution $\mathbf{Q}^{(l)}$ is further generated from $\mathbf{Q}^{(l)} \sim \mathcal{N}\left(\vec{q}, {\boldsymbol \sigma}^2\right)$. 


Formulation~\ref{eq:pdode1} is challenging to solve because the derivatives of the loss function with respect to $\vec{q}$ and ${\boldsymbol \sigma}$ are difficult to obtain. The reason is that $\mathbf{Q}^{(l)}$ is sampled from the Gaussian distribution, and it is difficult to compute $\frac{\partial \mathbf{Q}^{(l)}}{\partial \vec{q}}$ and $\frac{\partial \mathbf{Q}^{(l)}}{\partial {\boldsymbol \sigma}}$. Without the closed-form gradients, most existing studies adopt a two-step approach to estimate the PDOD. The first step estimates the OD demand mean and the second step estimates the standard deviation. The two steps are conducted iteratively until convergence \citep{ma2018statistical, yang2019estimating}. 
In this paper, we propose a novel solution to estimate the mean and standard deviation simultaneously by casting the PDODE problem into computational graphs. Details will be discussed in the following section.

\section{Solution Algorithm}
\label{sec:solution}
In this section, a reparameterization trick is developed to enable the simultaneous estimation of mean and standard deviation of the dynamic OD demand. The PDODE formulation in Equation~\ref{eq:pdode1} is then cast into a computational graph. We then summarize the step-by-step solution framework for PDODE. Finally, the underdetermination issue of the PDODE problem is discussed .

\subsection{A key reparameterization trick}

To solve Formulation~\ref{eq:pdode1}, our objective is to directly evaluate the derivative of both mean and standard deviation of OD demand, {\em i.e.} $\frac{\partial \mathcal{L}}{\partial \mathbf{q}}$ and $\frac{\partial \mathcal{L}}{\partial {\boldsymbol \sigma}}$, then gradient descent methods can be used to search for the optimal solution.

We will leave the computation of $\frac{\partial \mathcal{L}}{\partial \mathbf{q}}$ and $\frac{\partial \mathcal{L}}{\partial {\boldsymbol\sigma}}$ in the next section, while this section addresses a key issue of evaluating $\frac{\partial \vec{Q}^{(l)}}{\partial \mathbf{q}}$ and $\frac{\partial \vec{Q}^{(l)}}{\partial {\boldsymbol\sigma}}$. The idea is actually simple and straightforward. Instead of directly sampling $\vec{Q}^{(l)}$ from $\mathcal{N}\left(\vec{q}, {\boldsymbol\sigma}^2\right)$, we conduct the following steps to generate $\vec{Q}^{(l)}$: 1) Sample ${\boldsymbol\nu}^{(l)} \in \mathbb{R}^{NK}$ from $\mathcal{N}\left(\vec{0}, \mathbf{1}\right)$; 2) Obtain $\vec{Q}^{(l)}$ by $\vec{Q}^{(l)} = \vec{q} + {\boldsymbol\sigma}   \circ {\boldsymbol\nu}^{(l)}$, where $\circ $ represents the element-wise product. 

Through the above reparameterization trick, we can compute the derivatives $\frac{\partial \vec{Q}^{(l)}}{\partial \mathbf{q}}$ and $\frac{\partial \vec{Q}^{(l)}}{\partial {\boldsymbol\sigma}}$ by Equation~\ref{eq:odd}.
\begin{equation}
	\label{eq:odd}
	\begin{array}{llllll}
		\frac{\partial \vec{Q}^{(l)}}{\partial \mathbf{q}} &=& \vec{1}_{NK}\\
		\frac{\partial \vec{Q}^{(l)}}{\partial {\boldsymbol\sigma}} &=& {\boldsymbol\nu}^{(l)}
	\end{array}
\end{equation}
where $\vec{1}_{NK} \in \mathbb{R}^{NK}$ is a vector filled with $1$. This reparameterization trick is originally used to solve the variational autoencoder (VAE) \citep{kingma2013auto}, and we adapt it to solve the PDODE problem. 

\subsection{Reformulating PDODE through computational graphs}
With the reparameterization trick discussed in the previous section, we can reformulate the PDODE problem in Equation~\ref{eq:pdode2}.
\begin{equation}
	\label{eq:pdode2}
	\begin{array}{rrcllll}
		\vspace{5pt}
		\displaystyle \min_{\vec{q}, {\boldsymbol \sigma}} & \multicolumn{4}{l}{\displaystyle    \frac{1}{IL}\sum_{i=1}^I \sum_{l=1}^{L} \mathcal{M} \left(\tilde{\mathbf{X}}^{(i)}, \mathbf{X}^{(l)}\right)} &\\
		\textrm{s.t.} & \left \{ \vec{C}^{(l)}, \vec{T}^{(l)}, {\boldsymbol \rho}^{(l)} \ \right\} &= & \Lambda(\vec{F}^{(l)}) & \forall l&\\
		~ & \vec{p}^{(l)} &= & \Psi \left( \D(\vec{C}^{(l)}), \D(\vec{T}^{(l)})  \right)&  \forall l &\\
		~ & {\boldsymbol\nu}^{(l)} &\sim& \mathcal{N}\left(\vec{0}, \mathbf{1}\right)& \forall  l&\\
		~ & \mathbf{Q}^{(l)} & = & \vec{q} + {\boldsymbol\sigma}\circ{\boldsymbol\nu}^{(l)}& \forall  l&\\
		~ & \vec{F}^{(l)} & = & \vec{p}^{(l)}\vec{Q}^{(l)} & \forall l&\\
		~ & \mathbf{X}^{(l)} & = &  {\boldsymbol \rho}^{(l)} \vec{F}^{(l)} &\forall l &
	\end{array}
\end{equation}

Extending the forward-backward algorithm proposed by \citet{ma2019estimating}, we can solve Formulation~\ref{eq:pdode2} through the forward-backward algorithm. The forward-backward algorithm consists of two major components: 1) the forward iteration computes the objective function of Formulation~\ref{eq:pdode2}; 2) the backward iteration evaluates the gradients of the objective function with respect to the mean and standard deviation of the dynamic OD demand ($\frac{\partial \mathcal{L}}{\partial \mathbf{q}}, \frac{\partial \mathcal{L}}{\partial {\boldsymbol\sigma}}$). 

\begin{figure*}[h]
	\centering
	\includegraphics[width=0.95\linewidth]{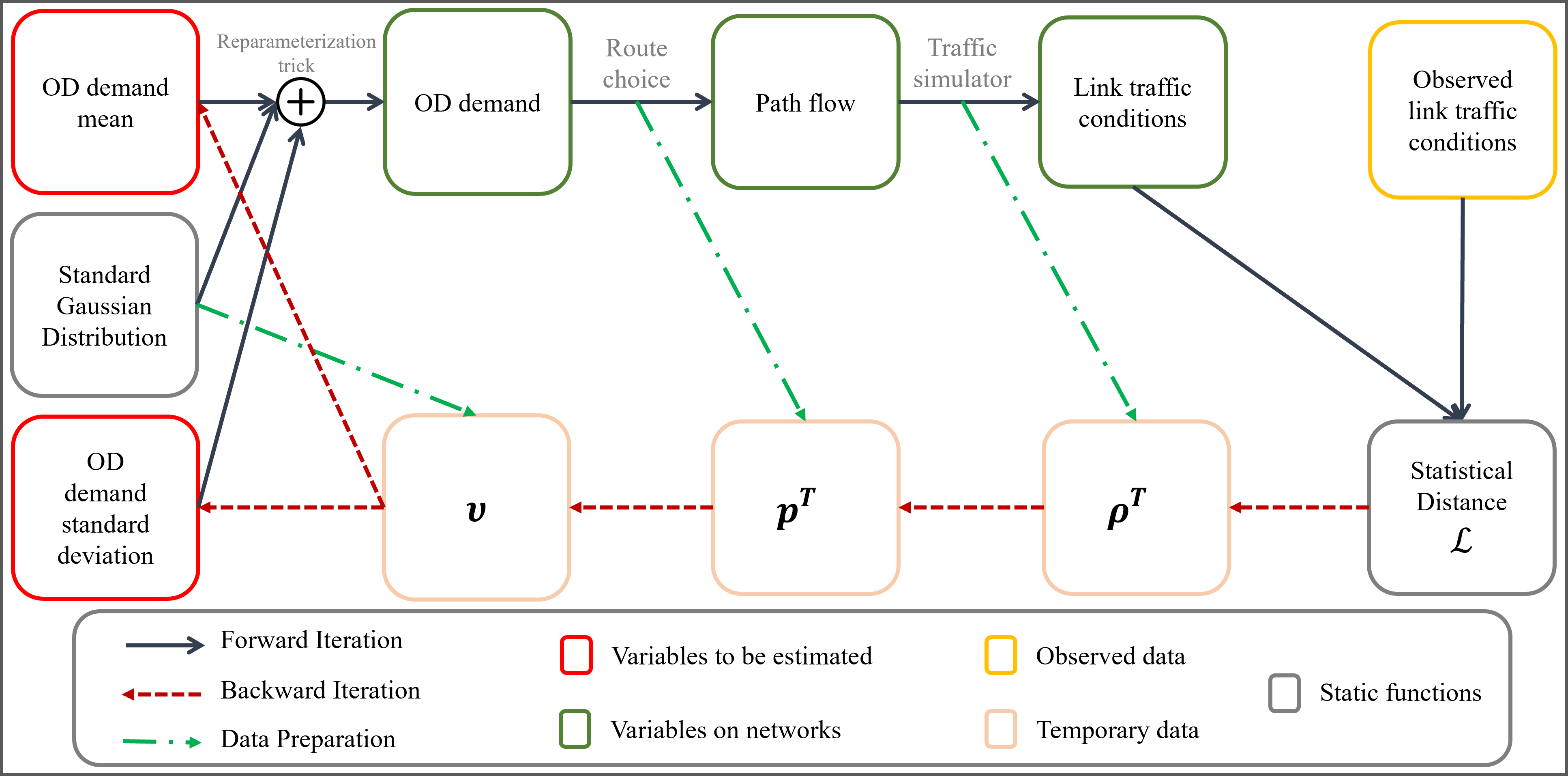}
	\caption{The computational graph for PDODE.}
	\label{fig:fb}
\end{figure*}

{\bf Forward iteration.} In the forward iteration, we compute the objective function based on the sample distribution of observation $\tilde{\mathbf{X}}^{(i)}$ in a decomposed manner, as presented in Equation~\ref{eq:forward}.
\begin{equation}
	\begin{array}{lllllll}
		\label{eq:forward}
		{\boldsymbol\nu}^{(l)} &\sim& \mathcal{N}\left(\vec{0}, \mathbf{1}\right)&\forall l\\
		\mathbf{Q}^{(l)} & = & \vec{q} + {\boldsymbol\sigma} {\boldsymbol \nu}^{(l)}&\forall l\\
		\vec{F}^{(l)} &=& \vec{p}^{(l)} \vec{Q}^{(l)}&\forall l\\
		\mathbf{X}^{(l)} &=& {\boldsymbol \rho}^{(l)}  \vec{F}^{(l)}&\forall l\\
		\mathcal{L} &=& \frac{1}{L}\sum_{l=1}^{L} \mathcal{M} \left(\tilde{\mathbf{X}}^{(i)}, \mathbf{X}^{(l)}\right)\\
	\end{array}
\end{equation}

{\bf Backward iteration.} The backward iteration evaluates the gradients of mean and standard deviation of the PDOD through the back-propagation (BP) method, as presented in Equation~\ref{eq:backward}.
\begin{equation}
	\begin{array}{llllllll}
		\label{eq:backward}
		\frac{\partial \mathcal{L}}{\partial \mathbf{X}^{(l)}} &=&  \frac{\partial \mathcal{M} \left(\tilde{\mathbf{X}}^{(i)}, \mathbf{X}^{(l)}\right)}{\partial \mathbf{X}^{(l)}} & \forall l\\
		\vspace{5pt}
		\frac{\partial \mathcal{L}}{\partial \vec{F}^{(l)}} &=& {{\boldsymbol \rho}^{(l)}}^T \frac{\partial \mathcal{L}}{\partial \mathbf{X}^{(l)}}& \forall l \\
		\vspace{5pt}
		\frac{\partial \mathcal{L}}{\partial \vec{Q}^{(l)}} &=& {\vec{p}^{(l)}}^T \frac{\partial \mathcal{L}}{\partial \vec{F}^{(l)}}& \forall l\\
		\vspace{5pt}
		\frac{\partial \mathcal{L}}{\partial \vec{q}} &= &\frac{\partial \mathcal{L}}{\partial \vec{Q}^{(l)}}& \forall l\\
		\vspace{5pt}
		\frac{\partial \mathcal{L}}{\partial {\boldsymbol\sigma}} &= & {\boldsymbol\nu}^{(l)}\circ \frac{\partial \mathcal{L}}{\partial \vec{Q}^{(l)}} & \forall l
	\end{array}
\end{equation}

The forward-backward algorithm is presented in Figure~\ref{fig:fb}. Forward iteration is conducted through the solid line, during which the temporary matrices ${\boldsymbol\nu}, \mathbf{p}, {\boldsymbol\rho}$ are also prepared (through the dot dashed link). The objective $\mathcal{L}$ is computed at the end of $\mathcal{L}$, and the backward iteration is conducted through the dashed line, and both OD demand mean and standard deviation are updated simultaneously.


In one iteration of forward-backward algorithm, we first run the forward iteration to compute the objective function, then the backward iteration is performed to evaluate the gradient of the objective function with respect to $\mathbf{q}, {\boldsymbol\sigma}$.
With the forward-backward algorithm to compute the gradient of the objective function, we can solve the PDODE formulation in Equation~\ref{eq:pdode2} through gradient-based methods. For example, the projected gradient descent method can be used to iteratively update the OD demand. This paper adopts Adagrad, a gradient-based method using adaptive step sizes \cite{duchi2011adaptive}. As for the stopping criteria, Proposition~\ref{prop:stop} indicates that the following two conditions are equivalent: 1) in the forward iteration, the distribution of path cost, link cost, path flow, OD demand do not change; 2) in the backward iteration, $\frac{\partial \mathcal{L}}{\partial \vec{q}} = 0$ and $\frac{\partial \mathcal{L}}{\partial {\boldsymbol\sigma}} = 0$. 

\begin{proposition}[Stopping criterion]
	\label{prop:stop}
	The PDODE formulation is solved when the forward and backward iterations converge, namely the the distributions of path cost, link cost, path flow, OD demand do not change, and $\frac{\partial \mathcal{L}}{\partial \vec{q}} = 0$ and $\frac{\partial \mathcal{L}}{\partial {\boldsymbol\sigma}} = 0$.
\end{proposition}

Since $\mathcal{L} \to \mathcal{L}_0$ when $I,L\to \infty$, we claim the PDODE problem is solved when $\frac{\partial \mathcal{L}}{\partial \vec{q}}$ and $\frac{\partial \mathcal{L}}{\partial {\boldsymbol\sigma}}$ are close to zero given a large $I$ and $L$.

\subsection{Solution Framework}
To summarize, the overall solution algorithm for PDODE is summarized in Table~\ref{tab:sol}.
\begin{table}[h]
	\begin{tabular}{p{2.2cm}p{13.6cm}}
		\textbf{Algorithm}& \textbf{[\textit{PDODE-FRAMEWORK}]} \\[3ex]\hline
		\textit{Step 0} & \textit{Initialization.} Initialize the mean and standard deviation of dynamic OD demand $\vec{q}, {\boldsymbol\sigma}$. \\[3ex]\hline
		\textit{Step 1} & \textit{Data preparation.} Randomly select a batch of observed data to form the sample distribution of $\tilde{\mathbf{X}}^{(i)}$.\\[3ex]\hline
		\textit{Step 2} & \textit{Forward iteration.}  Iterate over $l=1,\cdots,L$: for each $l$, sample ${\boldsymbol\nu}^{(l)}$, solve the DNL models and travelers' behavior model, and compute the objective function $\mathcal{L}$ based on Equation~\ref{eq:forward} with $\vec{q}, {\boldsymbol\sigma}$ \\[3ex]\hline
		\textit{Step 3} & \textit{Backward iteration.} Compute the gradient of the mean and standard deviation of dynamic OD demand using the backward iteration presented in Equation~\ref{eq:backward}. \\[3ex]\hline
		\textit{Step 4} & \textit{Update PDOD.}  Update the mean and standard deviation of dynamic OD ($\vec{q}, {\boldsymbol\sigma}$) with the gradient-based projection method.\\[3ex]\hline
		\textit{Step 5} & \textit{Batch Convergence check.}  Continue when the changes of OD demand mean and standard deviation are within tolerance. Otherwise, go to Step 2.\\[3ex]\hline
		\textit{Step 6} & \textit{Convergence check.}  Iterate over $i=1,\cdots, I$. Stop when the changes of OD demand mean and standard deviation are within tolerance across different $i$. Otherwise, go to Step 1.\\[3ex]\hline
	\end{tabular}
	\caption{The PDODE solution framework.}
	\label{tab:sol}
\end{table}

In practical applications, Step 3 and Step 4 can be conducted using the stochastic gradient projection methods to enhance the algorithm efficiency. 
Additionally, Step 3 and Step 4 can be conducted using the auto-differentiation and deep learning packages, such as PyTorch, TensorFlow, JAX, etc, and both steps can be run on multi-core CPUs and GPUs efficiently.

\subsection{Underdetermination and evaluation criterion}
In this section, we discuss the underdetermination issue for the PDODE problem. It is well known that both static OD estimation and dynamic OD estimation problems are undetermined \citep{yang1995heuristic,ma2018statistical}. We claim that PDODE is also under-determined because the problem dimension is much higher than its deterministic version. In the case of PDODE, not only the OD demand mean but also the standard deviation need to be estimated.
Therefore, estimating exact PDOD accurately with limited observed data is challenging in most practical applications. Instead, since the objective of PDODE is to better reproduce the observed traffic conditions, we can evaluate the PDODE methods based on whether they can reproduce the network conditions accurately. We can evaluate the PDODE framework by measuring how well the traffic conditions can be evaluated for the observed and all links, respectively. Using this concept, we categorize the PDODE evaluation criterion into three levels as follows:
\begin{enumerate}[label=\roman*)]
	\item Observed Links (\texttt{OL}): The traffic conditions simulated from the estimated PDOD on the observed links are accurate. 
	\item All Links (\texttt{AL}): The traffic conditions simulated from the estimated PDOD on all the links are accurate.
	\item Dynamic OD demand (\texttt{OD}): The estimated PDOD is accurate.
\end{enumerate}

\begin{figure*}[h]
	\centering
	\includegraphics[width=0.75\linewidth]{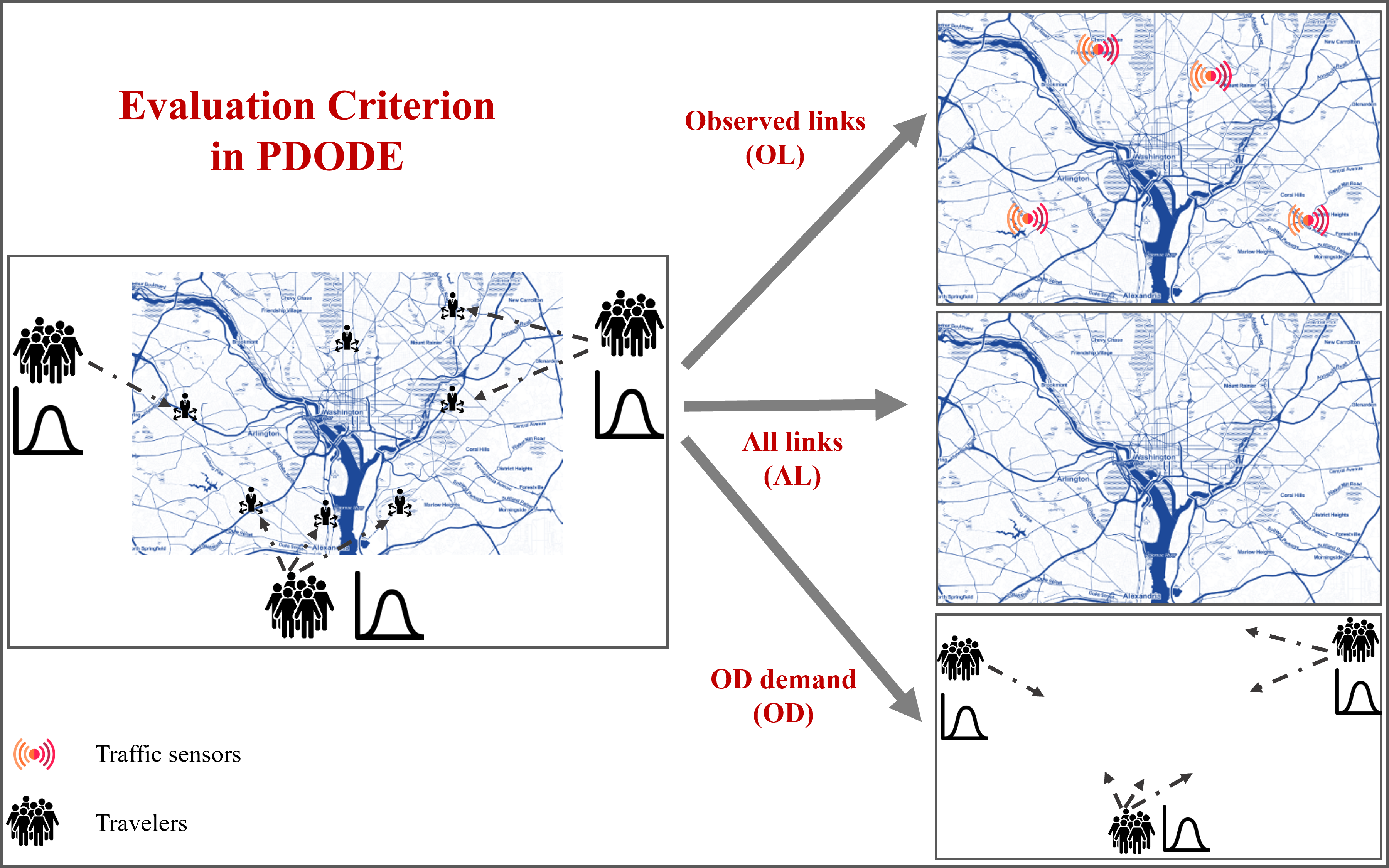}
	\caption{An overview of the evaluation criterion in PDODE.}
	\label{fig:ec}
\end{figure*}

The three evaluation criterion are summarized in Figure~\ref{fig:ec}. One can see that the objective of Formulation~\ref{eq:pdode2} is actually \texttt{OL}, and we include a series constraints in order to achieve \texttt{AL}. Specifically, the flow conservation and route choice model help to achieve \texttt{AL}. As for the \texttt{OD}, there is no guarantee for large-scale networks. Many recent studies also indicate the same observations \citep{osorio2019high, ma2019estimating, wollenstein2022joint}.

Overall, a PDODE framework that satisfies \texttt{OL} tends to overfit the observed data. We claim that a PDODE framework that satisfies \texttt{AL} is sufficient for most needs in traffic operation and management, as the ultimate goal for PDODE is to understand the dynamic traffic conditions on the networks. To achieve \texttt{OD}, a high-quality prior PDOD matrix is necessary to reduce the search space \citep{ma2018estimating}.

From the perspective of the underdetermination issue of PDODE, \texttt{OL} is always determined as it only focuses on the observed links. On general networks, \texttt{OD} is an under-determined problem as the cardinality of a network is much smaller than the dimension of OD demand. Whether \texttt{AL} is determined or not is based on the network topology and data availability, and hence it is promising to make full use of the proposed computational graphs to achieve \texttt{AL}, as the computational graphs have an advantage in multi-source data integration and fast computation.  This further motivates the necessity to formulate the PDODE problem using computational graphs.

\section{Numerical experiments}
\label{sec:experiment}
In this section, we first examine the proposed PDODE framework on a small network. Different statistical distances are compared and the optimal one is selected. We further compare the PDODE with DDODE method, and the parameter sensitivity is discussed. In addition, the effectiveness and scalability of the PDODE framework are demonstrated on a real-world large-scale network: SR-41 corridor. All the experiments in this section are conducted on a desktop with Intel Core i7-6700K CPU 4.00GHz $\times$ 8, 2133 MHz 2 $\times$ 16GB RAM, 500GB SSD.

\subsection{A small network}

\subsubsection{Settings}
\label{sec:setting}
\begin{figure}[h]
	\centering
	\includegraphics[width=0.75\linewidth]{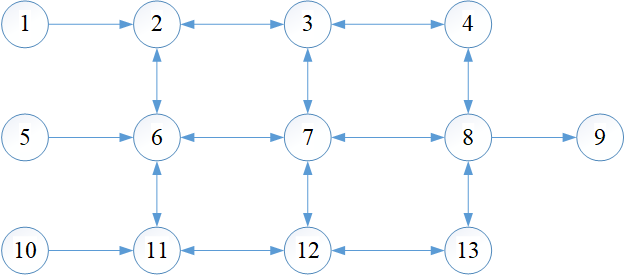}
	\caption{\footnotesize{An overview of the small network.}}
	\label{fig:31net}
\end{figure}

We first work on a small network with 13 nodes, 27 links, and 3 OD pairs, as presented in Figure~\ref{fig:31net}. There are in total 29 paths for the 3 OD pairs ($1 \to9$, $5 \to9$, and $10 \to9$). Links connecting node $1,5,9, 10$ are OD connectors, and the rest of links are standard roads with two lanes. The triangular fundamental diagrams (FD) are used for the standard links, in which the length of each road segment is $0.5$ mile, flow capacity is 2,000 vehicles/hour, and holding capacity is $200$ vehicles/mile. The free flow speed is uniform sampled from $20$ to $45$ miles/hour.


To evaluate the performance of the proposed method, we generate the mean and standard deviation of PDOD using a triangular pattern, as shown in Figure~\ref{fig:dod}. The PDOD is high enough to generate congestion. The observed flow is obtained by solving the statistical traffic equilibrium and then a Gaussian distributed noise is added to the observation. The performance of the PDODE estimation formulation is assessed by comparing the estimated flow
with the ``true'' flow (flow includes observed link flow, all link flow, and OD demand)  \citep{antoniou2015towards}. We set the study period to 10 time intervals and each time interval lasts 100 seconds. 
A Logit model with a dispersion factor $0.1$ is applied to the mean route travel time for modeling the travelers' behaviors. 

Supposing 100 days' data are collected, the dynamic OD demand from the ``true'' distribution of dynamic OD demand is generated on each day, and the demand is loaded on the network with the route choice model and DNL model. We randomly select 12 links to be observed, and a random Gaussian noise $\mathcal{N}(0, 5)$ is further added to the observed link flow. Our task is to estimate the mean and standard deviation of PDOD using the observed 100 days' data.

We run the proposed PDODE presented in Table~\ref{tab:sol} with the projected stochastic gradient descent, and the solution algorithm is Adagrad \citep{duchi2011adaptive}. We use the loss function $\mathcal{L}$ to measure the efficiency of the proposed method, as presented in Equation~\ref{eq:ereap}. 
Note that we loop all the 100 days' data once in each epoch.

\subsubsection{Comparing different statistical distances}

\begin{table*}[h]
	\centering
	\begin{tabular}{|l|cc|cc|cc|}
		\hline
		\multirow{2}{*}{\backslashbox{$\mathcal{M}$}{Accuracy}} & \multicolumn{2}{c|}{\texttt{OL}} & \multicolumn{2}{c|}{\texttt{AL}} & \multicolumn{2}{c|}{\texttt{OD}} \\
		\cline {2-7}
		~ & Mean & Std & Mean & Std & Mean & Std \\
		\hline\hline
		$\ell_1$-norm & {\bf 0.968}& 0.792& {\bf 0.997}& 0.806& {\bf 0.996}& 0.804\\
		$\ell_2$-norm & 0.955 & {\bf 0.880}& 0.994& {\bf 0.897} & 0.985& {\bf 0.892} \\
		2-Wasserstein distance & {\em 0.961} & {\em 0.843} & {\em 0.996} & {\em 0.861} & {\em 0.991}& {\em 0.860} \\
		KL divergence & -0.575 & 0.027 & 0.508 & 0.062 & -0.592 & 0.027 \\
		Bhattacharyya distance& -0.726 & -0.004 & 0.460 & 0.029 & -0.748 & -0.005 \\
		\hline
	\end{tabular}
	\caption{Performance of different statistical distances in terms of R-squared score.}
	\label{tab:compare}
\end{table*}

We first compare different statistical distances discussed in section~\ref{sec:obj}. Under the same settings in section~\ref{sec:setting}, different statistical distances are compared as the objective function in Formulation~\ref{eq:pdode2}, and the estimation results are presented in Table~\ref{tab:compare} in terms of R-squared score. We use the \texttt{r2\_score} function in sklearn to compute the R-squared score, and the score can be negative (because the model can be arbitrarily worse)\footnote{\url{https://scikit-learn.org/stable/modules/generated/sklearn.metrics.r2\_score.html}}.

One can see that neither KL divergence nor Bhattacharyya distance can yield proper estimation of PDOD, which may be due to the complicated formulations of its objective function, and gradient explosion and vanishing with respect to the objective function significantly affect the estimation accuracy. The other three statistical distances achieve satisfactory accuracy. Using $\ell_1$-norm and $\ell_2$-norm can achieve the best estimation of PDOD mean and standard deviation, respectively. Both objectives perform stably, which is probably attributed to the simple forms of their gradients. This finding is also consistent with many existing literature \citep{shao2014estimation,shao2015estimation}. The 2-Wasserstein distance achieves a balanced performance in terms of estimating both mean and standard deviation, which might be because the 2-Wasserstein distance compares the probability density functions, instead of directly comparing the parameters of the two distributions.  For the rest of the experiments, we choose 2-Wasserstein distance as the objective function.

\subsubsection{Basic estimation results}

\begin{figure}[h]
	\centering
	\includegraphics[width=0.95\linewidth]{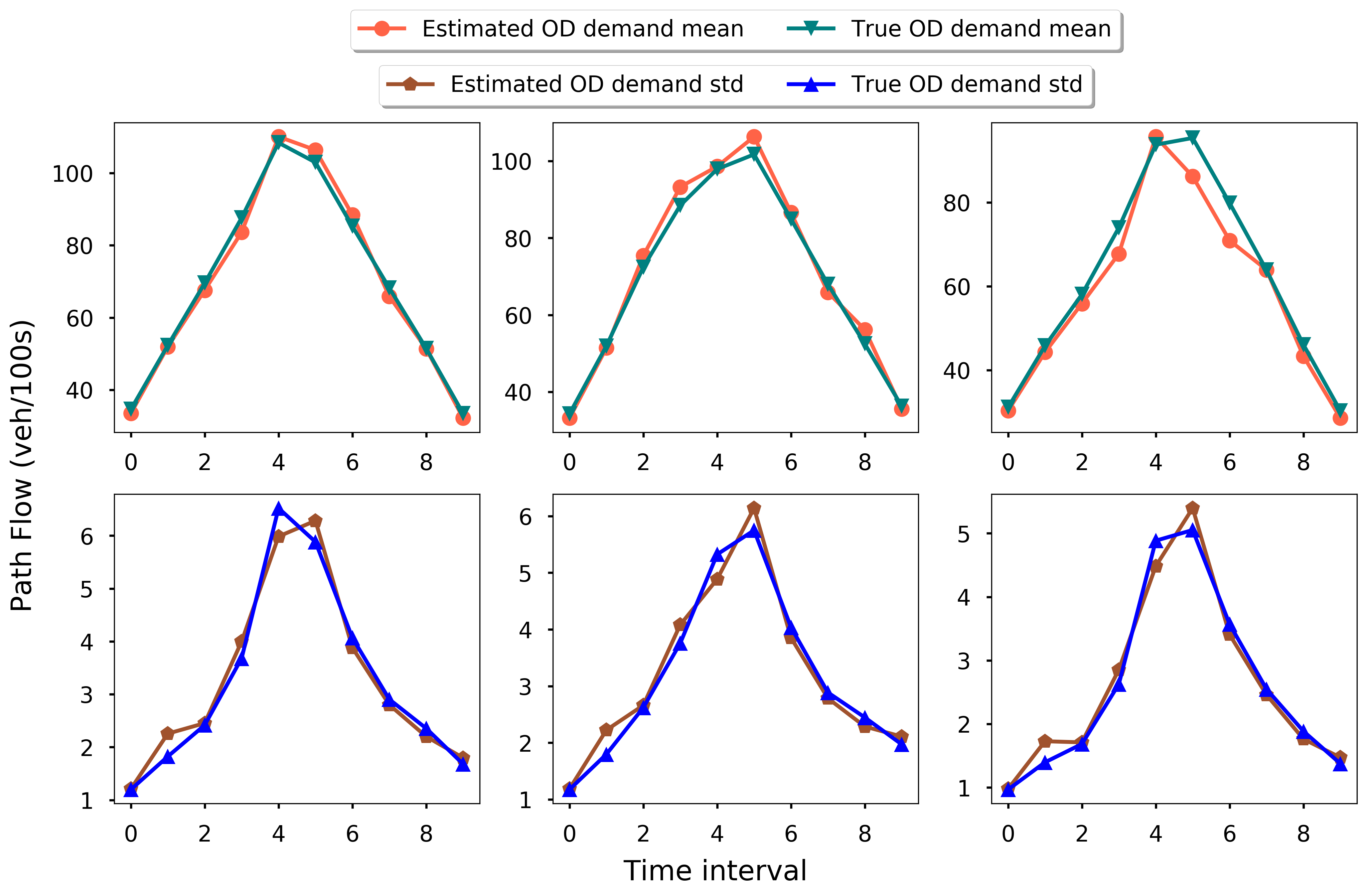}
	\caption{Comparison between the ``true'' and estimated OD demand (first row: mean; second row: standard deviation; first column: $1\to9$; second column: $5\to9$; third column: $10\to9$; unit: vehicle/100seconds).}
	\label{fig:dod}
\end{figure}

We present the detailed estimation results using 2-Wasserstein distance as the objective function. The estimated and ``true'' PDOD are compared in Figure~\ref{fig:dod}. One can see that the proposed PDODE framework accurately estimates the mean and standard deviation of the PDOD. Particularly, both surging and decreasing trends are quickly captured.

\subsubsection{Comparing with the deterministic DODE}

\begin{figure}[h]
	\centering
	\includegraphics[width=0.95\linewidth]{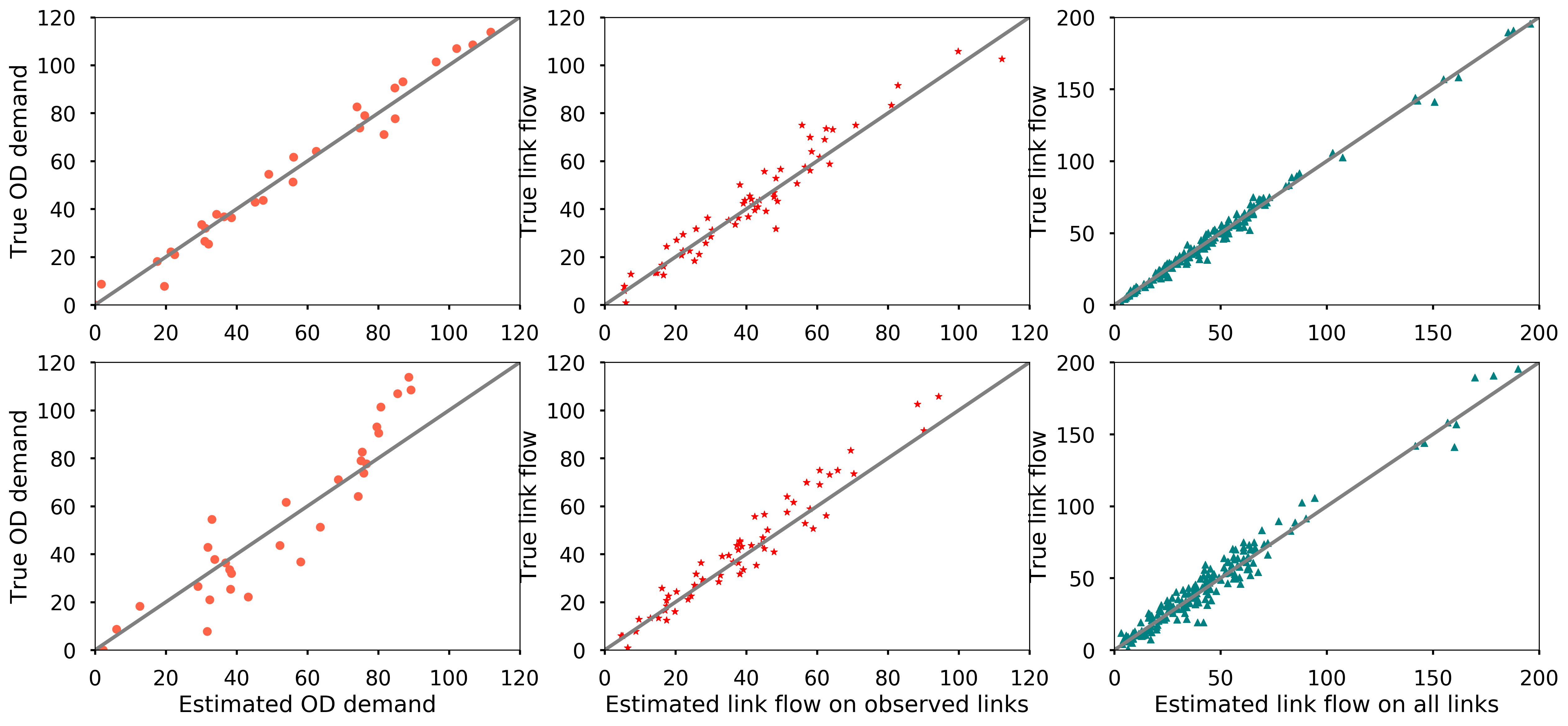}
	\caption{Comparison between the ``true'' and estimated flow in terms of \texttt{OL}, \texttt{AL}, and \texttt{OD} (first row: the proposed PDODE framework; second row: the standard DDODE framework; unit:vehicle/100seconds).}
	\label{fig:31comp}
\end{figure}

To demonstrate the advantages of the PDODE framework, we also run the standard DDODE framework proposed by \citet{ma2019estimating} using the same setting and data presented in section~\ref{sec:setting}. Because the DDODE framework does not estimate the standard deviation, so we only evaluate the estimation accuracy of the mean. The comparison is conducted by plotting the estimated OD demand mean, observed link flow and all link flow against ``true'' flow for both PDODE and DDODE frameworks, as presented in Figure~\ref{fig:31comp}. The algorithm performs well when the scattered points are close to $y=x$ line.

It can be seen from Figure~\ref{fig:31comp}, the PDODE framework can better reproduce the ``true'' traffic flow. Firstly,  DDODE can fit the observed link flow better as it directly optimize the gap between the observed and estimated link flow. However, the DDODE framework tends to overfit the noisy data because it does not model the variance of the flow explicitly. PDODE can provide a better estimation for those unobserved links and OD demand through a comprehensive modeling of the flow distribution. To summarize, DDODE achieves a higher accuracy on observed links (\texttt{OL}), while PDODE outperforms DDODE in terms of \texttt{AL} and \texttt{OD}.  

To quantify the error, we compute the R-squared scores between the ``true'' and estimated flow for both PDODE and DODE, as presented in Table~\ref{tab:sr}.

\begin{table}[h]
	\centering
	\begin{tabular}{|c|ccc|}
		\hline
		\backslashbox{Formulation}{Accuracy} & \texttt{OL} & \texttt{AL} & \texttt{OD} \\
		\hline\hline
		PDODE & 0.961 & {\bf 0.996} & {\bf 0.991} \\
		DDODE & {\bf 0.963} & 0.979 & 0.857 \\
		\hline
	\end{tabular}
	\caption{R-squared scores between the ``true'' and estimated flow for PDODE and DODE.}
	\label{tab:sr}
\end{table}

The R-squared score between the ``true'' and estimated OD demand and all links for PDODE is higher than that for DODE, while the differences between the R-squared scores for observed link flow are relatively small. This further explains the overfitting issue of the DDODE on the observed links, and the above experiments verify the illustrative example presented in section~\ref{sec:example}.

\subsubsection{Sensitivity analysis.} 

We also conduct sensitivity analysis regarding the proposed PDODE framework. 

{\bf Impact of travel time.}
If the travel time of each link on the network is also observed, the proposed PDODE framework can be extended to incorporate the data. To be specific, we use the travel time information to calibrate the DAR matrix using the approach presented in \citet{ma2018estimating}.
It is expected that the the estimation accuracy can further improved. The comparison of estimation accuracy is presented in Table~\ref{tab:compare2}. One can see that the inclusion of travel time data is beneficial to all the estimates (\texttt{OL}, \texttt{AL}, \texttt{OD}). Particularly, the estimation accuracy of standard deviation can be improved significantly by over 5\%.

\begin{table*}[h]
	\centering
	\begin{tabular}{|c|cc|cc|cc|}
		\hline
		\multirow{2}{*}{\backslashbox{$\mathcal{M}$}{Accuracy}} & \multicolumn{2}{c|}{\texttt{OL}} & \multicolumn{2}{c|}{\texttt{AL}} & \multicolumn{2}{c|}{\texttt{OD}} \\
		\cline {2-7}
		~ & Mean & Std & Mean & Std & Mean & Std \\
		\hline\hline
		PDODE & 0.961 & 0.843 & 0.996 & 0.861 & 0.991& 0.860 \\
		PDODE + travel time & 0.997& 0.925& 0.997& 0.921& 0.998& 0.908 \\
		\hline
		Improvement & +3.746\% & +9.727\% & +0.100\% & +9.969\% & +0.706\% & +5.581\%\\
		\hline
	\end{tabular}
	\caption{Effects of considering travel time data.}
	\label{tab:compare2}
\end{table*}

\begin{figure}[h]
	\centering
	\includegraphics[width=0.85\linewidth]{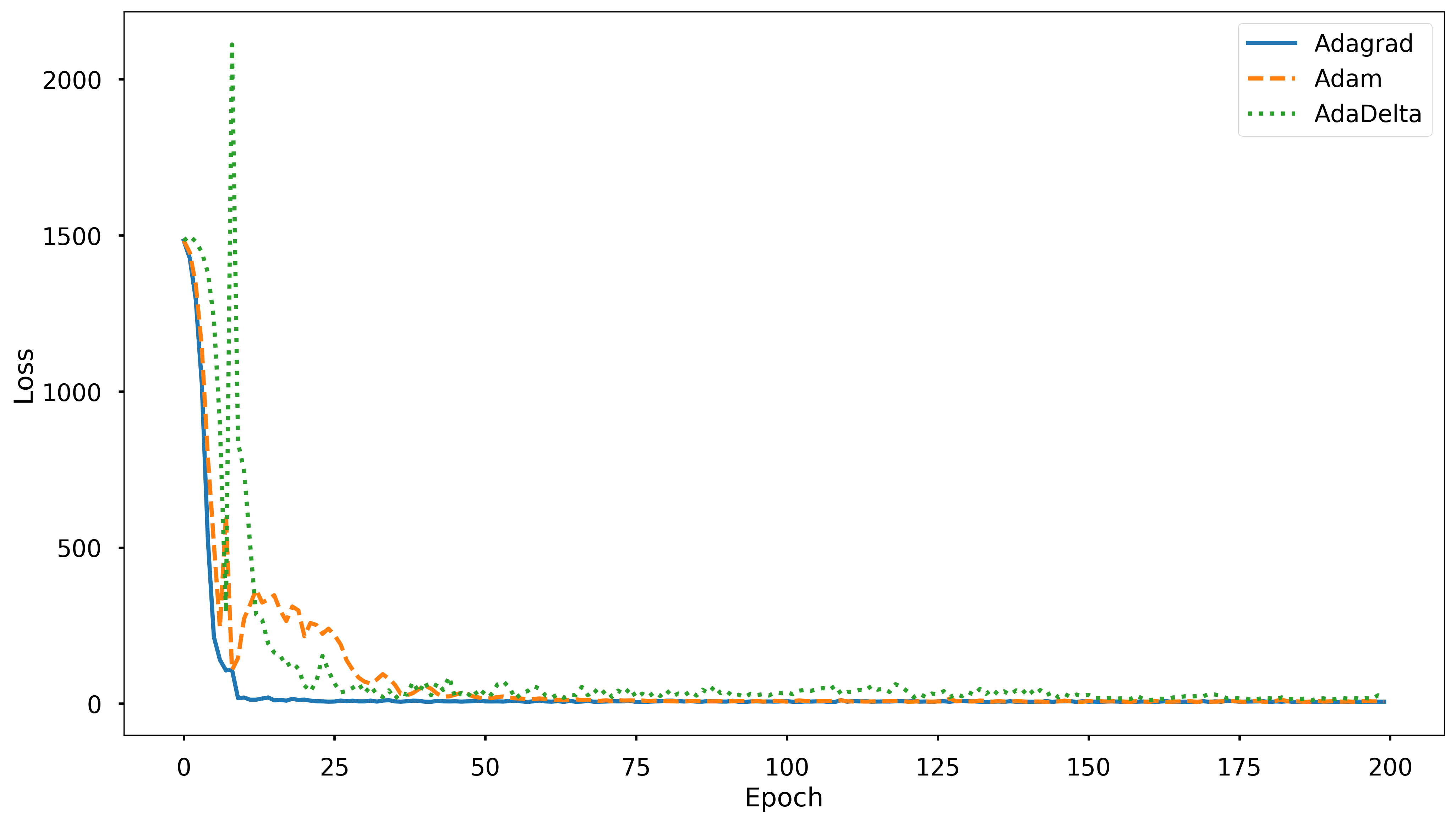}
	\caption{Convergence curves of different adaptive gradient descent methods.}
	\label{fig:conv}
\end{figure}

{\bf Adaptive gradient descent methods.}
We compare different adaptive gradient descent methods, and the convergence curves are shown in Figure~\ref{fig:conv}. Note that the conventional gradient descent (GD) or stochastic gradient descent (SGD) cannot converge and the loss does not decrease, so their curves are not shown in the figure. One can see that the Adagrad converges quickly within 20 epochs, and the whole 200 epochs take less than 10 minutes. Both Adam and AdaDelta can also converge after 60 epochs, while the curves are not as stable as the Adagrad. This is the reason we choose Adagrad in the proposed framework.

Sensitivity analysis regarding the learning rates, number of data samples, number of CPU cores, and noise level have also been conducted, and the results are similar to the previous study. For details, readers are referred to \citet{ma2019estimating}.

\subsection{A large-scale network: SR-41 corridor}
\begin{figure}[h]
	\centering
	\begin{subfigure}[b]{0.475\textwidth}
		\includegraphics[width=\textwidth]{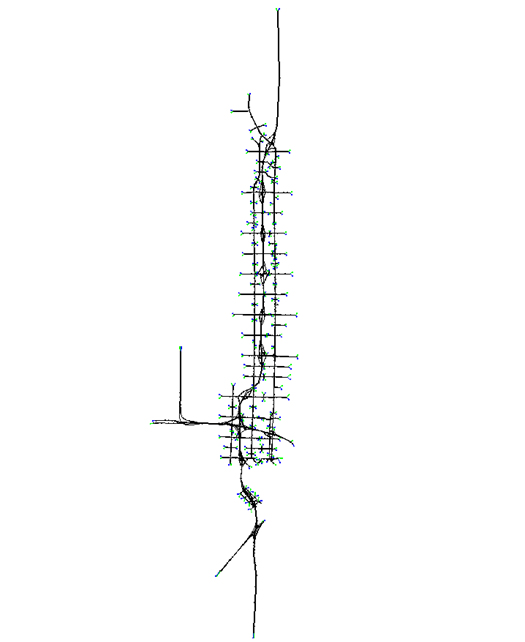}
		\caption{\footnotesize{Overview of the SR41 corridor.}}
		\label{fig:srnet}
	\end{subfigure}
	\begin{subfigure}[b]{0.475\textwidth}
		\includegraphics[width=\textwidth]{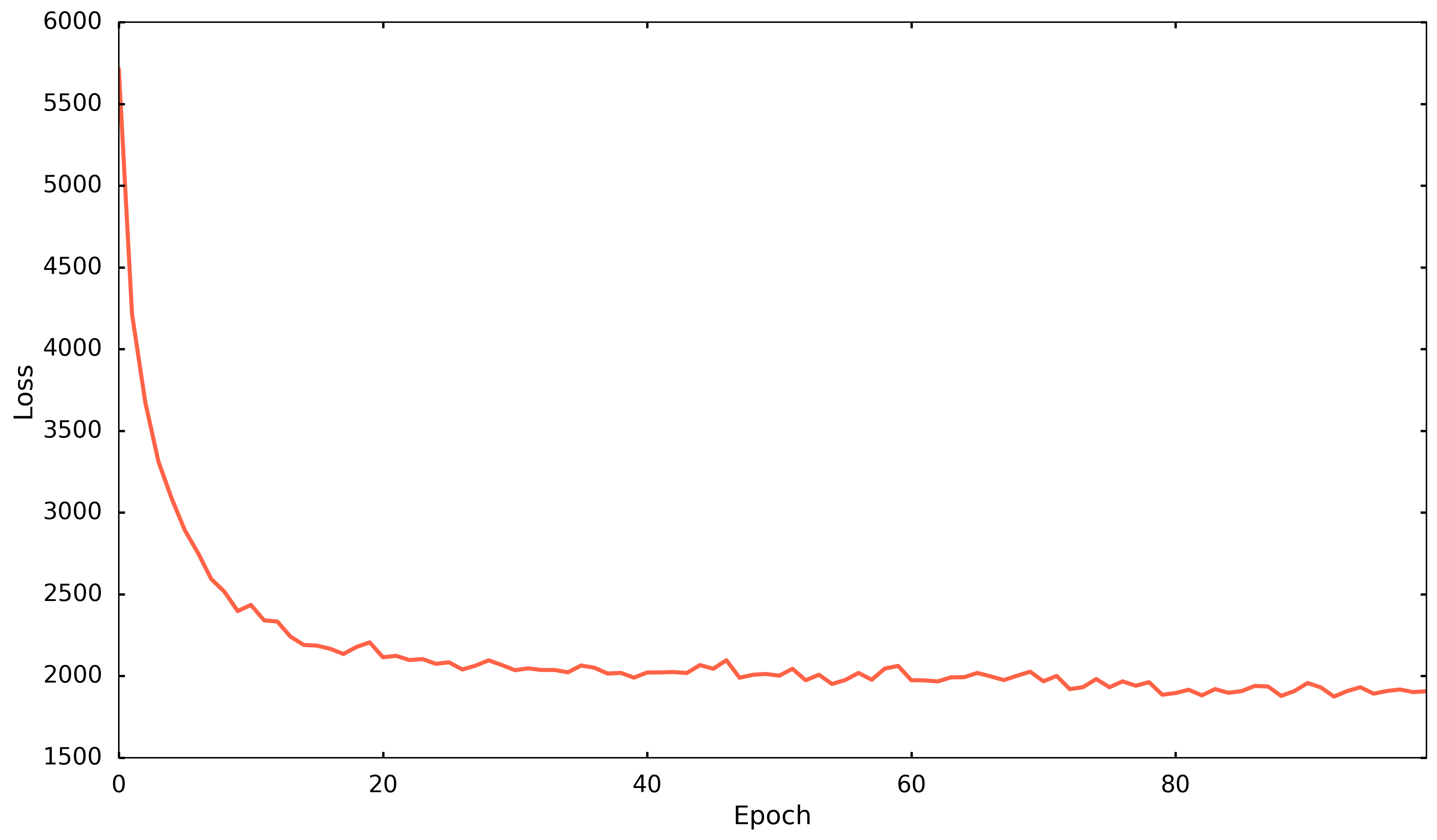}
		\caption{\footnotesize{Convergence curve of the proposed PDODE framework.}}
		\label{fig:convsr}
	\end{subfigure}
	\caption{Network overview and the algorithm convergence curve for SR41 corridor.}
	\label{fig:srover}
\end{figure}

In this section, we perform the proposed PDODE framework on a large-scale network. The SR-41 corridor is located in the City of Fresno, California. It consists of one major freeway and two parallel arterial roads. These roads are connected with local streets, as presented in Figure~\ref{fig:srnet}. The network contains 1,441 nodes, 2,413 links and 7,110 OD pairs \citep{liu2006streamlined, zhang2008developing}. We consider 6 time intervals and each time interval last 15 minutes. The ``true'' dynamic OD mean is generated from $\texttt{Unif}(0,5)$ and the standard deviation is generated from $\texttt{Unif}(0,1)$. It is assumed $500$ links are observed. The statistical traffic equilibrium is solved with Logit mode, and we generate $10$ days' data under the equilibrium. We run the proposed PDODE framework, and $\mathcal{L}$ with 2-Wasserstein distance is used to measure the efficiency of the algorithm. No historical OD demand is used, and the initial PDOD is randomly generated for the proposed solution algorithm. The convergence curve is presented in Figure~\ref{fig:convsr}. 

One can see the proposed framework performs well and the objective function converges quickly. Each epoch takes 37 minutes and hence the algorithm takes 3,700 minutes ($\sim$ 62 hours) to finish 100 epochs.

As discussed in previous sections, we do not compare the OD demand as the estimation of OD demand is under-determined, and it is challenging to fully recover the exact dynamic OD demand on such large-scale network without historical OD demand, analogous to DDODE or static ODE in the literature. Instead, we focus on the \texttt{OL} and \texttt{AL} to assess the performance of the proposed PDODE framework.

\begin{figure}[h]
	\centering
	\includegraphics[width=0.95\linewidth]{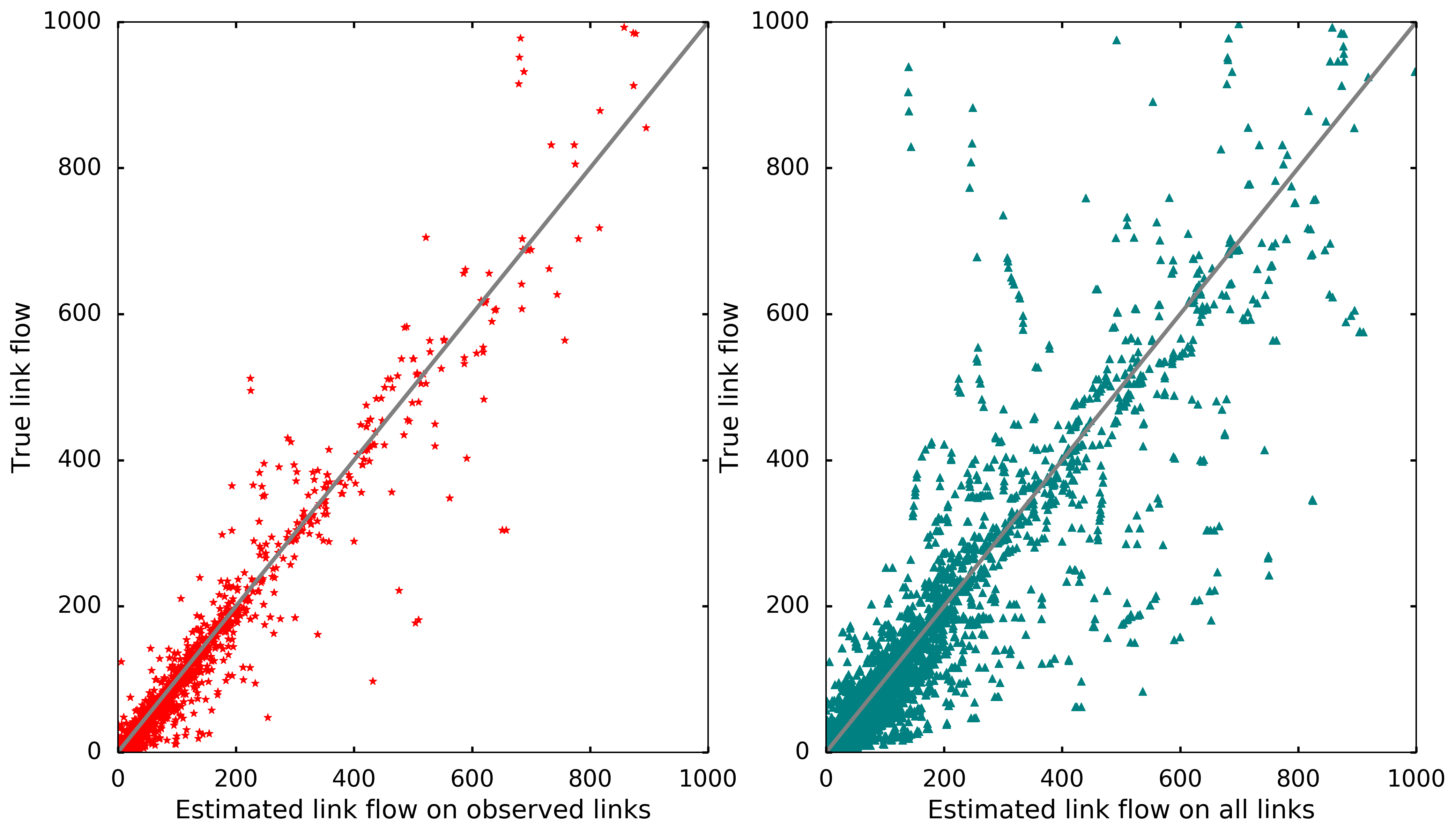}
	\caption{Comparison between the ``true'' and estimated mean of link flow in PDODE framework (unit:vehicle/15minutes).}
	\label{fig:srcomp}
\end{figure}

We plot the ``true'' and estimated mean of link flow on observed links and all links in Figure~\ref{fig:srcomp}, respectively. 
One can see that PDODE can reproduce the flow on observed links accurately, while the accuracy on all links is relatively lower. This observation is different from the small network, which implies extra difficulties and potential challenges of estimating dynamic network flow on large-scale networks, in extremely high dimensions. Quantitatively, the R-squared score is 0.949 on the observed links and 0.851 on all links. Both R-squared scores are satisfactory, and the R-squared score for \texttt{AL} is higher than that estimated by the DDODE framework ($0.823$). Hence we conclude that the proposed PDODE framework performs well on the large-scale network.

\section{Conclusions}
\label{sec:con}
This paper presents a data-driven framework for the probabilistic dynamic origin-destination demand estimation (PDODE) problem. The PDODE problem is rigorously formulated on general networks. 
Different statistical distances ({\em e.g.}, $\ell_p$-norm, Wasserstein distance, KL divergence, Bhattacharyya distance) are tested as the objective function. All the variables involved in the PDODE formulation are vectorized, and the proposed framework is cast into a computational graph.
Both mean and standard deviation of the PDOD can be simultaneously estimated through a novel reparameterization trick. The underdetermination issues of the PDODE problem are also discussed, and three different evaluation criterion (\texttt{OL}, \texttt{AL}, \texttt{OD}) are presented.

The proposed PDODE framework is examined on a small network as well as a real-world
large-scale network. The loss function reduces quickly on both networks and the time consumption is satisfactory. $\ell_1$ and $\ell_2$ norms have advantages in estimating the mean and standard deviation of dynamic OD demand, respectively, and the 2-Wasserstein distance achieves a balanced accuracy in estimating both mean and standard deviation. We also compared the DDODE framework with the proposed PDODE framework. The experiment results show that the DDODE framework tends to overfit on \texttt{OL}, and PDODE can achieve better estimation on \texttt{AL} and \texttt{OD}.

In the near future, we will extend the existing PDODE formulation to estimate the spatio-temporal covariance of the dynamic OD demand. The covariance (correlation) of dynamic OD demand can further help public agencies to better understand the intercorrelation of network dynamics and further improve the effectiveness of the operation/management strategies. Low-rank or sparsity regularization for the covariance matrix of the PDOD might be necessary. The choice of statistical distances can be better justified through theoretical derivations. The computational graph also has great potential in incorporating multi-source data \citep{ma2019estimating}, and it is interesting to explore the possibility of estimating the PDOD using emerging data sources, such as vehicle trajectory \citep{ma2019measuring} and automatic vehicle identification (AVI) data \citep{cao2021day}.  The under-determination issue remains a critical challenge for the OD estimation problem (including both DDODE and PDODE), and this study demonstrates the possibility of mitigating the overfitting issue by considering the standard deviation. We believe this sheds light on overcoming the under-determination issues in general OD estimation problems. 

\section*{Supplementary Materials}
The proposed PDODE framework is implemented with PyTorch and open-sourced on Github (\url{https://github.com/Lemma1/Probabilistic-OD-Estimation}).

\ACKNOWLEDGMENT{%
The work described in this paper was supported by U.S. National Science Foundation CMMI-1751448. The first author was  supported by the National Natural Science Foundation of China (No. 52102385) and a grant from the Research Grants Council of the Hong Kong Special Administrative Region, China (Project No. PolyU/25209221). The contents of this paper reflect the views of the authors, who are responsible for the facts and the accuracy of the information presented herein. 
}


%
%
%

\clearpage

\bibliographystyle{informs2014trsc}
\bibliography{ref}

\begin{thebibliography}{85}
\providecommand{\natexlab}[1]{#1}
\providecommand{\url}[1]{\texttt{#1}}
\providecommand{\urlprefix}{URL }

\bibitem[{Antoniou et~al.(2015{\natexlab{a}})Antoniou, Azevedo, Lu, Pereira,
  \protect\BIBand{} Ben-Akiva}]{antoniou2015w}
Antoniou C, Azevedo CL, Lu L, Pereira F, Ben-Akiva M, 2015{\natexlab{a}}
  \emph{W-spsa in practice: Approximation of weight matrices and calibration of
  traffic simulation models}. \emph{Transportation Research Part C: Emerging
  Technologies} 59:129--146.

\bibitem[{Antoniou et~al.(2015{\natexlab{b}})Antoniou, Barcel{\'o}, Breen,
  Bullejos, Casas, Cipriani, Ciuffo, Djukic, Hoogendoorn, Marzano
  et~al.}]{antoniou2015towards}
Antoniou C, Barcel{\'o} J, Breen M, Bullejos M, Casas J, Cipriani E, Ciuffo B,
  Djukic T, Hoogendoorn S, Marzano V, et~al., 2015{\natexlab{b}} \emph{Towards
  a generic benchmarking platform for origin--destination flows
  estimation/updating algorithms: Design, demonstration and validation}.
  \emph{Transportation Research Part C: Emerging Technologies} .

\bibitem[{Ashok \protect\BIBand{} Ben-Akiva(2000)}]{ashok2000alternative}
Ashok K, Ben-Akiva ME, 2000 \emph{Alternative approaches for real-time
  estimation and prediction of time-dependent origin--destination flows}.
  \emph{Transportation Science} 34(1):21--36.

\bibitem[{Bachir et~al.(2019)Bachir, Khodabandelou, Gauthier, El~Yacoubi,
  \protect\BIBand{} Puchinger}]{bachir2019inferring}
Bachir D, Khodabandelou G, Gauthier V, El~Yacoubi M, Puchinger J, 2019
  \emph{Inferring dynamic origin-destination flows by transport mode using
  mobile phone data}. \emph{Transportation Research Part C: Emerging
  Technologies} 101:254--275.

\bibitem[{Balakrishna, Ben-Akiva, \protect\BIBand{}
  Koutsopoulos(2008)}]{balakrishna2008time}
Balakrishna R, Ben-Akiva M, Koutsopoulos H, 2008 \emph{Time-dependent
  origin-destination estimation without assignment matrices}. \emph{Second
  International Symposium of Transport Simulation (ISTS06). Lausanne,
  Switzerland. 4-6 September 2006} (EPFL Press).

\bibitem[{Balmer et~al.(2009)Balmer, Rieser, Meister, Charypar, Lefebvre,
  \protect\BIBand{} Nagel}]{balmer2009matsim}
Balmer M, Rieser M, Meister K, Charypar D, Lefebvre N, Nagel K, 2009
  \emph{Matsim-t: Architecture and simulation times}. \emph{Multi-agent systems
  for traffic and transportation engineering}, 57--78 (IGI Global).

\bibitem[{Ben-Akiva et~al.(1998)Ben-Akiva, Bierlaire, Koutsopoulos,
  \protect\BIBand{} Mishalani}]{ben1998dynamit}
Ben-Akiva M, Bierlaire M, Koutsopoulos H, Mishalani R, 1998 \emph{Dynamit: a
  simulation-based system for traffic prediction}. \emph{DACCORD Short Term
  Forecasting Workshop}, 1--12.

\bibitem[{Ben-Akiva et~al.(2012)Ben-Akiva, Gao, Wei, \protect\BIBand{}
  Wen}]{ben2012dynamic}
Ben-Akiva ME, Gao S, Wei Z, Wen Y, 2012 \emph{A dynamic traffic assignment
  model for highly congested urban networks}. \emph{Transportation research
  part C: emerging technologies} 24:62--82.

\bibitem[{Bierlaire \protect\BIBand{} Crittin(2004)}]{bierlaire2004efficient}
Bierlaire M, Crittin F, 2004 \emph{An efficient algorithm for real-time
  estimation and prediction of dynamic od tables}. \emph{Operations Research}
  52(1):116--127.

\bibitem[{Bradbury et~al.(2018)Bradbury, Frostig, Hawkins, Johnson, Leary,
  Maclaurin, Necula, Paszke, Vander{P}las, Wanderman-{M}ilne, \protect\BIBand{}
  Zhang}]{jax2018github}
Bradbury J, Frostig R, Hawkins P, Johnson MJ, Leary C, Maclaurin D, Necula G,
  Paszke A, Vander{P}las J, Wanderman-{M}ilne S, Zhang Q, 2018 \emph{{JAX}:
  composable transformations of {P}ython+{N}um{P}y programs}.
  \urlprefix\url{http://github.com/google/jax}.

\bibitem[{Cao et~al.(2021)Cao, Tang, Sun, \protect\BIBand{} Ji}]{cao2021day}
Cao Y, Tang K, Sun J, Ji Y, 2021 \emph{Day-to-day dynamic origin--destination
  flow estimation using connected vehicle trajectories and automatic vehicle
  identification data}. \emph{Transportation Research Part C: Emerging
  Technologies} 129:103241.

\bibitem[{Cascetta, Inaudi, \protect\BIBand{}
  Marquis(1993)}]{cascetta1993dynamic}
Cascetta E, Inaudi D, Marquis G, 1993 \emph{Dynamic estimators of
  origin-destination matrices using traffic counts}. \emph{Transportation
  science} 27(4):363--373.

\bibitem[{Cipriani et~al.(2011)Cipriani, Florian, Mahut, \protect\BIBand{}
  Nigro}]{cipriani2011gradient}
Cipriani E, Florian M, Mahut M, Nigro M, 2011 \emph{A gradient approximation
  approach for adjusting temporal origin--destination matrices}.
  \emph{Transportation Research Part C: Emerging Technologies} 19(2):270--282.

\bibitem[{Cipriani et~al.(2021)Cipriani, Gemma, Mannini, Carrese,
  \protect\BIBand{} Crisalli}]{cipriani2021traffic}
Cipriani E, Gemma A, Mannini L, Carrese S, Crisalli U, 2021 \emph{Traffic
  demand estimation using path information from bluetooth data}.
  \emph{Transportation research part C: emerging technologies} 133:103443.

\bibitem[{Dantsuji et~al.(2022)Dantsuji, Hoang, Zheng, \protect\BIBand{}
  Vu}]{dantsuji2022novel}
Dantsuji T, Hoang NH, Zheng N, Vu HL, 2022 \emph{A novel metamodel-based
  framework for large-scale dynamic origin--destination demand calibration}.
  \emph{Transportation Research Part C: Emerging Technologies} 136:103545.

\bibitem[{Duchi, Hazan, \protect\BIBand{} Singer(2011)}]{duchi2011adaptive}
Duchi J, Hazan E, Singer Y, 2011 \emph{Adaptive subgradient methods for online
  learning and stochastic optimization.} \emph{Journal of machine learning
  research} 12(7).

\bibitem[{Fisk(1989)}]{fisk1989trip}
Fisk C, 1989 \emph{Trip matrix estimation from link traffic counts: the
  congested network case}. \emph{Transportation Research Part B:
  Methodological} 23(5):331--336.

\bibitem[{Florian \protect\BIBand{} Chen(1995)}]{florian1995coordinate}
Florian M, Chen Y, 1995 \emph{A coordinate descent method for the bi-level o--d
  matrix adjustment problem}. \emph{International Transactions in Operational
  Research} 2(2):165--179.

\bibitem[{Fl{\"o}tter{\"o}d(2017)}]{flotterod2017search}
Fl{\"o}tter{\"o}d G, 2017 \emph{A search acceleration method for optimization
  problems with transport simulation constraints}. \emph{Transportation
  Research Part B: Methodological} 98:239--260.

\bibitem[{Fl{\"o}tter{\"o}d, Bierlaire, \protect\BIBand{}
  Nagel(2011)}]{flotterod2011bayesian}
Fl{\"o}tter{\"o}d G, Bierlaire M, Nagel K, 2011 \emph{Bayesian demand
  calibration for dynamic traffic simulations}. \emph{Transportation Science}
  45(4):541--561.

\bibitem[{Han et~al.(2018)Han, Sun, Wang, \protect\BIBand{}
  Zhu}]{han2018stochastic}
Han L, Sun H, Wang DZ, Zhu C, 2018 \emph{A stochastic process traffic
  assignment model considering stochastic traffic demand}.
  \emph{Transportmetrica B: transport dynamics} 6(3):169--189.

\bibitem[{Hazelton(2008)}]{hazelton2008statistical}
Hazelton ML, 2008 \emph{Statistical inference for time varying
  origin--destination matrices}. \emph{Transportation Research Part B:
  Methodological} 42(6):542--552.

\bibitem[{Jha et~al.(2004)Jha, Gopalan, Garms, Mahanti, Toledo,
  \protect\BIBand{} Ben-Akiva}]{jha2004development}
Jha M, Gopalan G, Garms A, Mahanti B, Toledo T, Ben-Akiva M, 2004
  \emph{Development and calibration of a large-scale microscopic traffic
  simulation model}. \emph{Transportation Research Record: Journal of the
  Transportation Research Board} (1876):121--131.

\bibitem[{Jin \protect\BIBand{} Wen(2019)}]{jin2019behavior}
Jin L, Wen Y, 2019 \emph{Behavior and management of stochastic
  multiple-origin-destination traffic flows sharing a common link}. \emph{arXiv
  preprint arXiv:1903.05510} .

\bibitem[{Jin(2021)}]{jin2012link}
Jin WL, 2021 \emph{A link queue model of network traffic flow}.
  \emph{Transportation Science} 55(2):436--455.

\bibitem[{Kaack, Chen, \protect\BIBand{} Morgan(2019)}]{kaack2019truck}
Kaack LH, Chen GH, Morgan MG, 2019 \emph{Truck traffic monitoring with
  satellite images}. \emph{Proceedings of the Conference on Computing \&
  Sustainable Societies}, 155--164 (ACM).

\bibitem[{Kim, Zhou, \protect\BIBand{} Pendyala(2021)}]{kim2021computational}
Kim T, Zhou X, Pendyala RM, 2021 \emph{Computational graph-based framework for
  integrating econometric models and machine learning algorithms in emerging
  data-driven analytical environments}. \emph{Transportmetrica A: Transport
  Science} 1--30.

\bibitem[{Kingma \protect\BIBand{} Welling(2013)}]{kingma2013auto}
Kingma DP, Welling M, 2013 \emph{Auto-encoding variational bayes}. \emph{arXiv
  preprint arXiv:1312.6114} .

\bibitem[{LeBlanc \protect\BIBand{} Farhangian(1982)}]{leblanc1982selection}
LeBlanc LJ, Farhangian K, 1982 \emph{Selection of a trip table which reproduces
  observed link flows}. \emph{Transportation Research Part B: Methodological}
  16(2):83--88.

\bibitem[{Lee \protect\BIBand{} Ozbay(2009)}]{lee2009new}
Lee JB, Ozbay K, 2009 \emph{New calibration methodology for microscopic traffic
  simulation using enhanced simultaneous perturbation stochastic approximation
  approach}. \emph{Transportation Research Record: Journal of the
  Transportation Research Board} (2124):233--240.

\bibitem[{Liu et~al.(2006)Liu, Ding, Ban, Chen, \protect\BIBand{}
  Chootinan}]{liu2006streamlined}
Liu HX, Ding L, Ban JX, Chen A, Chootinan P, 2006 \emph{A streamlined network
  calibration procedure for california sr41 corridor traffic simulation study}.
  \emph{Proceedings of the 85th Transportation Research Board Annual Meeting}.

\bibitem[{Lu, Zhou, \protect\BIBand{} Zhang(2013)}]{lu2013dynamic}
Lu CC, Zhou X, Zhang K, 2013 \emph{Dynamic origin--destination demand flow
  estimation under congested traffic conditions}. \emph{Transportation Research
  Part C: Emerging Technologies} 34:16--37.

\bibitem[{Lu et~al.(2015)Lu, Xu, Antoniou, \protect\BIBand{}
  Ben-Akiva}]{lu2015enhanced}
Lu L, Xu Y, Antoniou C, Ben-Akiva M, 2015 \emph{An enhanced spsa algorithm for
  the calibration of dynamic traffic assignment models}. \emph{Transportation
  Research Part C: Emerging Technologies} 51:149--166.

\bibitem[{Ma, Zhang et~al.(2008)}]{ma2008polymorphic}
Ma J, Zhang HM, et~al., 2008 \emph{A polymorphic dynamic network loading
  model}. \emph{Computer-Aided Civil and Infrastructure Engineering}
  23(2):86--103.

\bibitem[{Ma, Ban, \protect\BIBand{} Pang(2014)}]{ma2014continuous}
Ma R, Ban XJ, Pang JS, 2014 \emph{Continuous-time dynamic system optimum for
  single-destination traffic networks with queue spillbacks}.
  \emph{Transportation Research Part B: Methodological} 68:98--122.

\bibitem[{Ma, Pi, \protect\BIBand{} Qian(2020)}]{ma2019estimating}
Ma W, Pi X, Qian S, 2020 \emph{Estimating multi-class dynamic
  origin-destination demand through a forward-backward algorithm on
  computational graphs}. \emph{Transportation Research Part C: Emerging
  Technologies} 119:102747.

\bibitem[{Ma \protect\BIBand{} Qian(2020)}]{ma2019measuring}
Ma W, Qian S, 2020 \emph{Measuring and reducing the disequilibrium levels of
  dynamic networks with ride-sourcing vehicle data}. \emph{Transportation
  Research Part C: Emerging Technologies} 110:222--246.

\bibitem[{Ma \protect\BIBand{} Qian(2017)}]{ma2017variance}
Ma W, Qian ZS, 2017 \emph{On the variance of recurrent traffic flow for
  statistical traffic assignment}. \emph{Transportation Research Part C:
  Emerging Technologies} 81:57--82.

\bibitem[{Ma \protect\BIBand{} Qian(2018{\natexlab{a}})}]{ma2018estimating}
Ma W, Qian ZS, 2018{\natexlab{a}} \emph{Estimating multi-year 24/7
  origin-destination demand using high-granular multi-source traffic data}.
  \emph{Transportation Research Part C: Emerging Technologies} 96:96--121.

\bibitem[{Ma \protect\BIBand{} Qian(2018{\natexlab{b}})}]{ma2018statistical}
Ma W, Qian ZS, 2018{\natexlab{b}} \emph{Statistical inference of probabilistic
  origin-destination demand using day-to-day traffic data}.
  \emph{Transportation Research Part C: Emerging Technologies} 88:227--256.

\bibitem[{Mahmassani \protect\BIBand{} Herman(1984)}]{mahmassani1984dynamic}
Mahmassani H, Herman R, 1984 \emph{Dynamic user equilibrium departure time and
  route choice on idealized traffic arterials}. \emph{Transportation Science}
  18(4):362--384.

\bibitem[{Mahmassani, Hu, \protect\BIBand{}
  Jayakrishnan(1992)}]{mahmassani1992dynamic}
Mahmassani H, Hu T, Jayakrishnan R, 1992 \emph{Dynamic traffic assignment and
  simulation for advanced network informatics (dynasmart)}. \emph{Proceedings
  of the 2nd international CAPRI seminar on Urban Traffic Networks, Capri,
  Italy}.

\bibitem[{Mahmassani et~al.(2014)Mahmassani, Kim, Chen, Stogios, Brijmohan,
  \protect\BIBand{} Vovsha}]{mahmassani2014incorporating}
Mahmassani HS, Kim J, Chen Y, Stogios Y, Brijmohan A, Vovsha P, 2014
  \emph{Incorporating reliability performance measures into operations and
  planning modeling tools} (Transportation Research Board).

\bibitem[{Nakayama(2016)}]{nakayama2016effect}
Nakayama S, 2016 \emph{Effect of providing traffic information estimated by a
  stochastic network equilibrium model with stochastic demand}.
  \emph{Transportation Research Part C: Emerging Technologies} 70:238--251.

\bibitem[{Nakayama \protect\BIBand{} Watling(2014)}]{nakayama2014consistent}
Nakayama S, Watling D, 2014 \emph{Consistent formulation of network equilibrium
  with stochastic flows}. \emph{Transportation Research Part B: Methodological}
  66:50--69.

\bibitem[{Nguyen(1977)}]{nguyen1977estimating}
Nguyen S, 1977 \emph{Estimating and OD Matrix from Network Data: a Network
  Equilibrium Approach} (Montr{\'e}al: Universit{\'e} de Montr{\'e}al, Centre
  de recherche sur les transports).

\bibitem[{Nie \protect\BIBand{} Zhang(2008)}]{nie2008variational}
Nie YM, Zhang HM, 2008 \emph{A variational inequality formulation for inferring
  dynamic origin--destination travel demands}. \emph{Transportation Research
  Part B: Methodological} 42(7):635--662.

\bibitem[{Nie \protect\BIBand{} Zhang(2010)}]{nie2010solving}
Nie YM, Zhang HM, 2010 \emph{Solving the dynamic user optimal assignment
  problem considering queue spillback}. \emph{Networks and Spatial Economics}
  10(1):49--71.

\bibitem[{Oh et~al.(2019)Oh, Seshadri, Azevedo, \protect\BIBand{}
  Ben-Akiva}]{oh2019demand}
Oh S, Seshadri R, Azevedo CL, Ben-Akiva ME, 2019 \emph{Demand calibration of
  multimodal microscopic traffic simulation using weighted discrete spsa}.
  \emph{Transportation Research Record} 0361198119842107.

\bibitem[{Osorio(2019{\natexlab{a}})}]{osorio2019dynamic}
Osorio C, 2019{\natexlab{a}} \emph{Dynamic origin-destination matrix
  calibration for large-scale network simulators}. \emph{Transportation
  Research Part C: Emerging Technologies} 98:186--206.

\bibitem[{Osorio(2019{\natexlab{b}})}]{osorio2019high}
Osorio C, 2019{\natexlab{b}} \emph{High-dimensional offline origin-destination
  (od) demand calibration for stochastic traffic simulators of large-scale road
  networks}. \emph{Transportation Research Part B: Methodological} 124:18--43.

\bibitem[{Patil et~al.(2022)Patil, Behara, Khadhir, \protect\BIBand{}
  Bhaskar}]{patil2022methods}
Patil SN, Behara KN, Khadhir A, Bhaskar A, 2022 \emph{Methods to enhance the
  quality of bi-level origin--destination matrix adjustment process}.
  \emph{Transportation Letters} 1--10.

\bibitem[{Patwary, Huang, \protect\BIBand{} Lo(2021)}]{patwary2021metamodel}
Patwary AU, Huang W, Lo HK, 2021 \emph{Metamodel-based calibration of
  large-scale multimodal microscopic traffic simulation}. \emph{Transportation
  Research Part C: Emerging Technologies} 124:102859.

\bibitem[{Pi, Ma, \protect\BIBand{} Qian(2018)}]{CARTRUCK}
Pi X, Ma W, Qian S, 2018 \emph{Regional traffic planning and operation using
  mesoscopic car-truck traffic simulation}.

\bibitem[{Qian(2016)}]{qian2016dynamic}
Qian S, 2016 \emph{Dynamic network analysis \& real-time traffic management for
  philadelphia metropolitan area} .

\bibitem[{Qian, Shen, \protect\BIBand{} Zhang(2012)}]{qian2012system}
Qian ZS, Shen W, Zhang H, 2012 \emph{System-optimal dynamic traffic assignment
  with and without queue spillback: Its path-based formulation and solution via
  approximate path marginal cost}. \emph{Transportation research part B:
  methodological} 46(7):874--893.

\bibitem[{Qurashi et~al.(2019)Qurashi, Ma, Chaniotakis, \protect\BIBand{}
  Antoniou}]{qurashi2019pc}
Qurashi M, Ma T, Chaniotakis E, Antoniou C, 2019 \emph{Pc-spsa: Employing
  dimensionality reduction to limit spsa search noise in dta model
  calibration}. \emph{IEEE Transactions on Intelligent Transportation Systems}
  .

\bibitem[{Ros-Roca et~al.(2022)Ros-Roca, Montero, Barcel{\'o}, N{\"o}kel,
  \protect\BIBand{} Gentile}]{ros2022practical}
Ros-Roca X, Montero L, Barcel{\'o} J, N{\"o}kel K, Gentile G, 2022 \emph{A
  practical approach to assignment-free dynamic origin--destination matrix
  estimation problem}. \emph{Transportation Research Part C: Emerging
  Technologies} 134:103477.

\bibitem[{Shao et~al.(2014)Shao, Lam, Sumalee, Chen, \protect\BIBand{}
  Hazelton}]{shao2014estimation}
Shao H, Lam WH, Sumalee A, Chen A, Hazelton ML, 2014 \emph{Estimation of mean
  and covariance of peak hour origin--destination demands from day-to-day
  traffic counts}. \emph{Transportation Research Part B: Methodological}
  68:52--75.

\bibitem[{Shao et~al.(2015)Shao, Lam, Sumalee, \protect\BIBand{}
  Hazelton}]{shao2015estimation}
Shao H, Lam WH, Sumalee A, Hazelton ML, 2015 \emph{Estimation of mean and
  covariance of stochastic multi-class od demands from classified traffic
  counts}. \emph{Transportation Research Part C: Emerging Technologies} .

\bibitem[{Shao, Lam, \protect\BIBand{} Tam(2006)}]{shao2006reliability}
Shao H, Lam WH, Tam ML, 2006 \emph{A reliability-based stochastic traffic
  assignment model for network with multiple user classes under uncertainty in
  demand}. \emph{Networks and Spatial Economics} 6(3-4):173--204.

\bibitem[{Shen et~al.(2019)Shen, Shao, Wu, \protect\BIBand{}
  Lam}]{shen2019spatial}
Shen L, Shao H, Wu T, Lam WH, 2019 \emph{Spatial and temporal analyses for
  estimation of origin-destination demands by time of day over year}.
  \emph{IEEE Access} 7:47904--47917.

\bibitem[{Shen, Nie, \protect\BIBand{} Zhang(2007)}]{shen2007path}
Shen W, Nie Y, Zhang HM, 2007 \emph{On path marginal cost analysis and its
  relation to dynamic system-optimal traffic assignment}. \emph{Transportation
  and Traffic Theory 2007. Papers Selected for Presentation at ISTTT17}.

\bibitem[{Sheppard et~al.(2017)Sheppard, Waraich, Campbell, Pozdnukov,
  \protect\BIBand{} Gopal}]{sheppard2017modeling}
Sheppard C, Waraich R, Campbell A, Pozdnukov A, Gopal AR, 2017 \emph{Modeling
  plug-in electric vehicle charging demand with beam: the framework for
  behavior energy autonomy mobility}. Technical report, Lawrence Berkeley
  National Lab.(LBNL), Berkeley, CA (United States).

\bibitem[{Speelpenning(1980)}]{speelpenning1980compiling}
Speelpenning B, 1980 \emph{Compiling fast partial derivatives of functions
  given by algorithms}. Ph.D. thesis, University of Illinois at
  Urbana-Champaign.

\bibitem[{Srivastava et~al.(2014)Srivastava, Hinton, Krizhevsky, Sutskever,
  \protect\BIBand{} Salakhutdinov}]{srivastava2014dropout}
Srivastava N, Hinton G, Krizhevsky A, Sutskever I, Salakhutdinov R, 2014
  \emph{Dropout: a simple way to prevent neural networks from overfitting}.
  \emph{The journal of machine learning research} 15(1):1929--1958.

\bibitem[{Stolte, Tang, \protect\BIBand{} Hanrahan(2002)}]{stolte2002polaris}
Stolte C, Tang D, Hanrahan P, 2002 \emph{Polaris: A system for query, analysis,
  and visualization of multidimensional relational databases}. \emph{IEEE
  Transactions on Visualization and Computer Graphics} 8(1):52--65.

\bibitem[{Sun et~al.(2019)Sun, Guo, Wu, Zhu, Wu, Xian, \protect\BIBand{}
  Zhou}]{sun2019analyzing}
Sun J, Guo J, Wu X, Zhu Q, Wu D, Xian K, Zhou X, 2019 \emph{Analyzing the
  impact of traffic congestion mitigation: From an explainable neural network
  learning framework to marginal effect analyses}. \emph{Sensors} 19(10):2254.

\bibitem[{Tavana(2001)}]{tavana2001internally}
Tavana H, 2001 \emph{Internally-consistent estimation of dynamic network
  origin-destination flows from intelligent transportation systems data using
  bi-level optimization} .

\bibitem[{Tympakianaki, Koutsopoulos, \protect\BIBand{}
  Jenelius(2015)}]{tympakianaki2015c}
Tympakianaki A, Koutsopoulos HN, Jenelius E, 2015 \emph{c-spsa: Cluster-wise
  simultaneous perturbation stochastic approximation algorithm and its
  application to dynamic origin--destination matrix estimation}.
  \emph{Transportation Research Part C: Emerging Technologies} 55:231--245.

\bibitem[{Vaze et~al.(2009)Vaze, Antoniou, Wen, \protect\BIBand{}
  Ben-Akiva}]{vaze2009calibration}
Vaze V, Antoniou C, Wen Y, Ben-Akiva M, 2009 \emph{Calibration of dynamic
  traffic assignment models with point-to-point traffic surveillance}.
  \emph{Transportation Research Record: Journal of the Transportation Research
  Board} (2090):1--9.

\bibitem[{Waller et~al.(2021)Waller, Chand, Zlojutro, Nair, Niu, Wang, Zhang,
  \protect\BIBand{} Dixit}]{waller2021rapidex}
Waller ST, Chand S, Zlojutro A, Nair D, Niu C, Wang J, Zhang X, Dixit VV, 2021
  \emph{Rapidex: A novel tool to estimate origin--destination trips using
  pervasive traffic data}. \emph{Sustainability} 13(20):11171.

\bibitem[{Watling et~al.(2015)Watling, Rasmussen, Prato, \protect\BIBand{}
  Nielsen}]{watling2015stochastic}
Watling DP, Rasmussen TK, Prato CG, Nielsen OA, 2015 \emph{Stochastic user
  equilibrium with equilibrated choice sets: Part i--model formulations under
  alternative distributions and restrictions}. \emph{Transportation Research
  Part B: Methodological} 77:166--181.

\bibitem[{Wollenstein-Betech et~al.(2022)Wollenstein-Betech, Sun, Zhang,
  Cassandras, \protect\BIBand{} Paschalidis}]{wollenstein2022joint}
Wollenstein-Betech S, Sun C, Zhang J, Cassandras CG, Paschalidis IC, 2022
  \emph{Joint data-driven estimation of origin-destination demand and travel
  latency functions in multi-class transportation networks}. \emph{IEEE
  Transactions on Control of Network Systems} .

\bibitem[{Wu et~al.(2018)Wu, Guo, Xian, \protect\BIBand{}
  Zhou}]{wu2018hierarchical}
Wu X, Guo J, Xian K, Zhou X, 2018 \emph{Hierarchical travel demand estimation
  using multiple data sources: A forward and backward propagation algorithmic
  framework on a layered computational graph}. \emph{Transportation Research
  Part C: Emerging Technologies} 96:321--346.

\bibitem[{Yang(1995)}]{yang1995heuristic}
Yang H, 1995 \emph{Heuristic algorithms for the bilevel origin-destination
  matrix estimation problem}. \emph{Transportation Research Part B:
  Methodological} 29(4):231--242.

\bibitem[{Yang et~al.(1992)Yang, Sasaki, Iida, \protect\BIBand{}
  Asakura}]{yang1992estimation}
Yang H, Sasaki T, Iida Y, Asakura Y, 1992 \emph{Estimation of
  origin-destination matrices from link traffic counts on congested networks}.
  \emph{Transportation Research Part B: Methodological} 26(6):417--434.

\bibitem[{Yang, Fan, \protect\BIBand{} Royset(2019)}]{yang2019estimating}
Yang Y, Fan Y, Royset JO, 2019 \emph{Estimating probability distributions of
  travel demand on a congested network}. \emph{Transportation Research Part B:
  Methodological} 122:265--286.

\bibitem[{Yu et~al.(2021)Yu, Zhu, Yang, Guo, \protect\BIBand{}
  Tang}]{yu2021bayesian}
Yu H, Zhu S, Yang J, Guo Y, Tang T, 2021 \emph{A bayesian method for dynamic
  origin--destination demand estimation synthesizing multiple sources of data}.
  \emph{Sensors} 21(15):4971.

\bibitem[{Zhang et~al.(2008)Zhang, Ma, Singh, \protect\BIBand{}
  Chu}]{zhang2008developing}
Zhang H, Ma J, Singh SP, Chu L, 2008 \emph{Developing calibration tools for
  microscopic traffic simulation final report part iii: Global calibration-od
  estimation, traffic signal enhancements and a case study}. \emph{PATH Rep.
  UCB-ITS-PRR-2008} 8.

\bibitem[{Zhang, Nie, \protect\BIBand{} Qian(2013)}]{zhang2013modelling}
Zhang H, Nie Y, Qian Z, 2013 \emph{Modelling network flow with and without link
  interactions: the cases of point queue, spatial queue and cell transmission
  model}. \emph{Transportmetrica B: Transport Dynamics} 1(1):33--51.

\bibitem[{Zhang et~al.(2021)Zhang, Chen, Yang, Ma, Jin, \protect\BIBand{}
  Gao}]{zhang2021network}
Zhang J, Chen F, Yang L, Ma W, Jin G, Gao Z, 2021 \emph{Network-wide link
  travel time and station waiting time estimation using automatic fare
  collection data: A computational graph approach}. \emph{arXiv preprint
  arXiv:2108.09292} .

\bibitem[{Zhou \protect\BIBand{} Mahmassani(2007)}]{zhou2007structural}
Zhou X, Mahmassani HS, 2007 \emph{A structural state space model for real-time
  traffic origin--destination demand estimation and prediction in a day-to-day
  learning framework}. \emph{Transportation Research Part B: Methodological}
  41(8):823--840.

\bibitem[{Zhou \protect\BIBand{} Taylor(2014)}]{zhou2014dtalite}
Zhou X, Taylor J, 2014 \emph{Dtalite: A queue-based mesoscopic traffic
  simulator for fast model evaluation and calibration}. \emph{Cogent
  Engineering} 1(1):961345.

\bibitem[{Zinkevich, Langford, \protect\BIBand{}
  Smola(2009)}]{zinkevich2009slow}
Zinkevich M, Langford J, Smola AJ, 2009 \emph{Slow learners are fast}.
  \emph{Advances in neural information processing systems}, 2331--2339.

\end{thebibliography}




\end{document}